\documentclass[preprint]{aastex631}

\shorttitle{$e$ and $\beta$ distribution of the disrupted stars}
\shortauthors{Zhong, Hayasaki, Li, Berczik \& Spurzem}
\graphicspath{{./}{figures/}}

\usepackage{amsmath}
\usepackage{amssymb}
\usepackage{graphicx,graphics}
\usepackage[english]{babel}
\usepackage{color}

\begin{document}

\title{Exploring the origin of stars on bound and unbound orbits causing tidal disruption events}

\author[0000-0003-4121-5684]{Shiyan Zhong}
\affiliation{South-Western Institute for Astronomy Research, Yunnan University, Kunming, 650500 Yunnan, China}

\author[0000-0003-4799-1895]{Kimitake Hayasaki}
\affiliation{Department of Astronomy and Space Science, Chungbuk National University, Cheongju 361-763, Korea}
\affiliation{Harvard-Smithsonian Center for Astrophysics, 60 Garden Street, Cambridge, MA02138, USA}

\author[0000-0001-6530-0424]{Shuo Li}
\affiliation{National Astronomical Observatories of China, Chinese Academy of Sciences, 20A Datun Rd., Chaoyang District, 100012, Beijing, China}


\author[0000-0003-4176-152X]{Peter Berczik}
\affiliation{Astronomisches Rechen-Institut, Zentrum f\"ur Astronomie, University of Heidelberg, M\"onchhofstrasse 12-14, 69120, Heidelberg, Germany}
\affiliation{Nicolaus Copernicus Astronomical Centre Polish Academy of Sciences, ul. Bartycka 18, 00-716 Warsaw, Poland}
\affiliation{Konkoly Observatory, Research Centre for Astronomy and Earth Sciences, E\"otv\"os Lor\'and Research Network (ELKH), MTA Centre of Excellence, Konkoly Thege Mikl\'os \'ut 15-17, 1121 Budapest, Hungary}
\affiliation{Main Astronomical Observatory, National Academy of Sciences of Ukraine, 27 Akademika Zabolotnoho St., 03680, Kyiv, Ukraine}


\author[0000-0003-2264-7203]{Rainer Spurzem}
\affiliation{Astronomisches Rechen-Institut, Zentrum f\"ur Astronomie, University of Heidelberg, M\"onchhofstrasse 12-14, 69120, Heidelberg, Germany}
\affiliation{Kavli Institute for Astronomy and Astrophysics, Peking University, Yiheyuan Lu 5, Haidian Qu, 100871, Beijing, China}
\affiliation{National Astronomical Observatories of China, Chinese Academy of Sciences, 20A Datun Rd., Chaoyang District, 100012, Beijing, China}

\def\red#1{\textcolor{red}{\textbf{#1}}}
\def\blue#1{\textcolor{blue}{\textbf{#1}}}
\def\black#1{\textcolor{black}{\textbf{#1}}}
\def\mag#1{\textcolor{magenta}{\textbf{#1}}}
\def\green#1{\textcolor{blue}{\textbf{#1}}}
\def\vec#1{\mbox{\boldmath $#1$}}

\begin{abstract}

Tidal disruption events (TDEs) provide a clue to the properties of a central supermassive black hole (SMBH) and an accretion disk around it, and to the stellar density and velocity distributions in the nuclear star cluster surrounding the SMBH. Deviations of TDE light curves from the standard occurring at a parabolic encounter with the SMBH depends on whether the stellar orbit is hyperbolic or eccentric (Hayasaki et al. 2018) and the penetration factor ($\beta$, tidal disruption radius to orbital pericenter ratio). We study the orbital parameters of bound and unbound stars being tidally disrupted by comparison of direct $N$-body simulation data with an analytical model. Starting from the classical steady-state Fokker-Planck model of \cite{CK1978}, we develop an analytical model of the number density distribution of those stars as a function of orbital eccentricity ($e$) and $\beta$. To do so fittings of the density and velocity distribution of the nuclear star cluster and of the energy distribution of tidally disrupted stars are required and obtained from $N$-body data.
We confirm that most of the stars causing TDEs in a spherical nuclear star cluster originate from the full loss-cone region of phase space, derive analytical boundaries in eccentricity-$\beta$ space, and find them confirmed by $N$-body data. Since our limiting eccentricities are much smaller than critical eccentricities for full accretion or full escape of stellar debris, we conclude that those stars are only very marginally eccentric or hyperbolic, close to parabolic.

\end{abstract}


\section{Introduction}

Most galaxies harbor supermassive black holes (SMBHs) with millions to billions of solar masses at their center. Tidal disruption events (TDEs) provide a good probe to identify dormant SMBHs in inactive galaxies. 
A star is tidally disrupted by an SMBH when the star approaches the SMBH closely enough that the black hole's tidal force exceeds the stellar self-gravity \citep{H1975}.
In classical TDE theory, a star on a parabolic orbit and is 
tidally disrupted by the SMBH at tidal disruption radius,
$r_{\rm t}=(M_{\rm BH}/m_*)^{1/3}r_*$, where $M_{\rm BH}$, $m_*$, and $r_*$
are the black hole mass, stellar mass and radius, respectively. 
Subsequently, half of the stellar debris falls back to the SMBH at a rate of $t^{-5/3}$ so that the bolometric luminosity is proportional to $t^{-5/3}$ if the mass fallback rate equals the mass accretion rate \citep{R1988,EK1989,1989IAUS..136..543P}.
However, recent observations have revealed
that some observed TDEs show light curves, which deviate from the $t^{-5/3}$
decay rate \citep{GCR2012,HPB2014,2015ApJ...815L...5G,MKM2015,HKP2016,2019ApJ...878...82V}.
Dozens of X-ray TDEs have light curves shallower than $t^{-5/3}$ \citep{2017ApJ...838..149A},
while many optical/UV TDEs are well fit by $t^{-5/3}$ (e.g. \citet{2017ApJ...842...29H}).

Some possible reasons for the deviation of the light curve from the $t^{-5/3}$ law are discussed in current literature.
The following are the three main reasons among them. First, the fallback debris would cause a self-crossing shock by a relativistic apsidal precession \citep{2015ApJ...804...85S,2015ApJ...806..164P,2020ApJ...904...73R}, outflowing a significant fraction of the debris \cite{2020MNRAS.492..686L}. 
Moreover, the secondary shock due to subsequently occurring collision forms an accretion disk. The bolometric luminosity at the photosphere clearly deviates from the $t^{-5/3}$ decay \citep{2020MNRAS.495.1374B}. 
Second, even though the mass fallback rate follows the $t^{-5/3}$ law, the radiative fluxes emitted from the accretion disk or disk wind modify the light curve variation.
\citet{LR2011} have shown that the luminosity of the accretion disk observed in different
bands may decay with diverse power-law indexes. For TDEs with well-evolved optically thick accretion disks, the observed X-ray light curves should decay following the form of a power-law
multiplied by an exponential, which is caused by the Wien tail of the disk spectrum
\citep{2020MNRAS.492.5655M}.
Moreover, when mass falls back to the SMBH at a super-Eddington rate, an outflow could be launched
from the accretion disk. The luminosity of the outflow can decay with time shallower than
$t^{-5/3}$ \citep{SQ2009}.
Final, the mass fallback rate can deviate from the standard $t^{-5/3}$ rate due to external and internal properties of the tidally disrupted star, such as: its
orbital eccentricity \citep{HSL2013,HZL2018,2020ApJ...900....3P}; its
orbital energy and angular momentum, the combination of these two quantities defines how deep its orbit
reaches inside the tidal radius (we define a penetration factor $\beta=r_{\rm t}/r_{\rm p}$, where $r_{\rm p}$ is the pericenter of the
stellar orbit around the black hole);
the detailed stellar internal structure and the possible survival of a stellar core during a partial tidal disruption \citep{GRR2013}.

In this paper, we focus on the number density distribution of the stars, which cause TDEs, as a function of orbital eccentricity and the penetration factor.
Since these orbital parameters leave imprints on the observable flux, the $e$ and $\beta$ could be obtained by fitting the light curves of the observed TDE with, e.g., \texttt{MOSFiT},~\citet{MOSFiT}. We can presume the number density by using this observed $e$ and $\beta$ in the end.
Currently about a dozen of TDEs are known with $\beta$ measurements \citep{MGR2019,NBB2019,GNS2020}, but they do not get the eccentricity independently, because the \texttt{MOSFiT} software used is based on hydrodynamic simulations of \citet{GRR2013}, who only simulated the $e=1$ case.
We note from this analysis that all the measured values of $\beta$ are close to unity.
\citet{SM2016} suggested that the $\beta$ value of the stars originating from the empty loss-cone regime should be very close to unity, while for the full loss-cone regime $\beta$ could take larger values and the number density is proportional to $\beta^{-2}$ (the definitions of empty and full loss-cone are given in Section~\ref{Theory-CK78}).
As we will discuss in Section~\ref{sub:dis}, also
in the full loss-cone regime many orbits may have $\beta\sim 1$; larger $\beta$ values occur in this case, but with a smaller probability. The eccentricity could provide further constraints on this issue, but as indicated  before the eccentricity was set to exactly unity for all measurements so far.

In a preceding paper \citep{HZL2018} we have examined the distribution of tidally disrupted stars on the $e$-$\beta$ plane by using $N$-body experiments of spherical nuclear star clusters. We found two interesting results: first, the eccentricities of the stars that causes TDEs usually take values between two critical eccentricities proposed by \cite{2020ApJ...900....3P}, which depend on $\beta$, and so there is some correlation between $e$ and $\beta$. Second, the distributions of $e$ and $\beta$ vary with the mass ratio between stars and the central SMBH, as do the critical eccentricities. This is important for future studies using stars of different masses - here and in \cite{HZL2018} we just limit ourselves to stars of equal mass. Light curve characteristics of tidal disruption flares depend on the eccentricity and its critical limits. This raises further interest in the distribution of orbital parameters.

In this paper, we analytically derive the number densities of bound and unbound stars that undergo TDEs in a spherical nuclear star cluster and test them by comparison with $N$-body simulations.
We also examine the distribution of stars on the $e$-$\beta$ plane by estimating the allowed eccentricity range for a given $\beta$.
Predicting the relative frequency of TDEs with different eccentricity $e$ and penetration factor $\beta$ should help identifying realistic values of $e$ and $\beta$ for future hydrodynamic simulations of TDEs and also the interpretation of TDE observations (constraining the dynamic processes operating in the host cluster).
We construct our analytical models in Section~\ref{Theory} and then describe the details of $N$-body simulations and compare the analytical number densities to the simulation results in Section~\ref{SECT-TEST}. We discuss our results in Section~\ref{SECT-DISCUSSION} and draw our conclusions in Section~\ref{sec:cons}.

\section{Orbital parameter dependence of the stellar distribution}
\label{Theory}
%
In this section we first briefly review the theory of the steady-state stellar distribution around a central black hole in a star cluster of \cite{CK1978} (hereafter CK78). It originates from a numerical solution of the orbit-averaged Fokker-Planck equation in energy-angular momentum space. The CK78 solution describes the distribution of stars inside the cusp surrounding the central SMBH, assuming the gravitational potential is dominated by the SMBH. It is well suited as a starting point to derive the stellar distribution in a phase space of orbital eccentricity $e$ and penetration factor $\beta$ (see Section~\ref{Theory-bound-star}). These quantities are of interest here, because they provide relevant input parameters for hydrodynamic simulations of TDEs, which in turn provide key information about the nature of TDE light curves. For our analysis of star cluster simulations with TDEs it is an advantage to use parameters closely related to the TDE and its observational characteristics, rather than the more conventional integrals of stellar motion.

In Section~\ref{Theory-unbound-star} we discuss the case of unbound stars experiencing TDEs. Earlier work by \cite{MT1999,WM2004} and \cite{SM2016} is based on a generalized treatment following CK78, using the Fokker-Planck equation also in the region of a galaxy unbound to the SMBH. For our purposes we choose a simpler but still useful approach in that regime. In what follows, we use subscript `b' and `u' to mark the quantities corresponding to bound and unbound cases, respectively.

%
\subsection{Stellar distributions in energy-angular momentum space}
\label{Theory-CK78}
%
The influence radius $r_{\rm h}$ of an SMBH in the
center of a star cluster is defined as the radius within which the enclosed stellar mass equals to the SMBH mass. Inside $r_{\rm h}$ stars are considered as gravitationally bound to the black hole and a stellar density cusp forms.
Following CK78 we characterize a stellar orbit in this region by both the specific orbital energy of a star
$E = v^2/2 - GM_{\rm BH}/r$ and by its normalized squared angular momentum
\begin{eqnarray}
\mathcal{R} = \left(\frac{J}{J_{\rm c}}\right)^2,
\label{eq:angmom}
\end{eqnarray}
where $v$, $G$, $J$ and $J_{\rm c}$ are the velocity of the star, gravitation constant, specific angular momentum of the star and the corresponding circular angular momentum, respectively. In a spherically symmetric cusp with isotropic velocity dispersion, the stellar density distribution depends on the orbital energy only.\footnote{Note that spherical symmetry does not necessarily imply isotropy, and TDE rates and properties in strongly anisotropic systems may differ substantially from the standard isotropic case \citep{MW2005,LV2015,SGV2018}.} However, the assumption of isotropic velocity distribution breaks down because the SMBH removes stars with low angular momentum through TDEs. So, in classical loss-cone theory, the stellar number density $n$ should also depend on $\mathcal{R}$, and thus be a function of both $E$ and $\mathcal{R}$ i.e. $n = n(\mathcal{R}, E)$ \citep{CK1978}. In phase space the loss-cone region is encompassed by $\mathcal{R}_{\rm lc}$, where $\mathcal{R}_{\rm lc}$ is the square of the normalized loss-cone angular momentum (see equation~\ref{Eq-R_lc}). For the models described in this paper stars inside the loss-cone region can survive for no more than one orbital period, unless they find a way out before being disrupted (typically by being scattered out of the loss-cone by two-body relaxation). In reality partial tidal disruptions may occur \citep{Zhongetal2022,McLeodetal2013}, stars may not be fully disrupted at the first passage near the tidal radius. In case of only full tidal disruptions considered in this paper the loss-cone runs out of stars quickly and $n(\mathcal{R},E)$ vanishes to zero at $\mathcal{R} \ll \mathcal{R}_{\rm lc}$. Simplified models based on moment equations of the Fokker-Planck equation \citep{2001MNRAS.327..995A,2004MNRAS.352..655A} assumed a sudden drop of $n$ to zero at $\mathcal{R}_{\rm lc}$, while the original work of CK78 shows the solution around and inside $\mathcal{R}_{\rm lc}$ follows a logarithmic profile and reaches zero at $\mathcal{R}_0$ (defined by equation~\ref{Eq-R0}).
Two-body relaxation encounters replenish the loss-cone by angular momentum diffusion. So, in steady state
$n(\mathcal{R},E)$ is determined by  
an equilibrium between disruption processes near the tidal radius and the replenishment process. 
By solution of the orbit averaged Fokker-Planck equation, taking both processes into account, CK78 found the following expression for the stellar density:
\begin{equation}
n_{\rm CK}(\mathcal{R},E) \simeq A(E) \ln \left( \frac{\mathcal{R}}{\mathcal{R}_{\rm 0}}
\right)\hspace{0.5cm}(\mathcal{R} > \mathcal{R}_{\rm 0}),
\label{Eq-f(R,E)}
\end{equation}
\noindent
where $A(E)$ is an energy-dependent coefficient and $\mathcal{R}_{\rm 0}$ is the square of the normalized angular momentum at the zero-boundary below which the number density goes to zero.
The CK78 solution was limited to the Keplerian potential. Later works, such as \citet{MT1999} and \citet{WM2004} that have taken the stellar potential into account also reported the logarithmic dependence on $\mathcal{R}$.

Our focus is on the bound stars that cause TDEs
(i.e. $\mathcal{R}_0 \leq \mathcal{R} \leq \mathcal{R}_{\rm lc}$), the cumulative number density, $N_{\rm TD,b}(\mathcal{R},E)$, has the same $\ln\mathcal{R}$-dependence as equation~\ref{Eq-f(R,E)} (because they originate from the stellar cusp described by the CK78 distribution) but the normalization coefficient is different,

\begin{equation}
N_{\rm TD,b}(\mathcal{R},E) = A_{\rm TD,b}(E) \ln \left( \frac{\mathcal{R}}{\mathcal{R}_{\rm 0}}
\right)\hspace{0.5cm}(\mathcal{R}_0 \leq \mathcal{R} \leq \mathcal{R}_{\rm lc}),
\label{Eq-n_TD(R,E)}
\end{equation}
\noindent
where the new coefficient $A_{\rm TD,b}(E)$ is obtained by the normalization $N_{\rm TD,b}(E) = \int_{\mathcal{R}_0}^{\mathcal{R}_{\rm lc}} N_{\rm TD,b}(\mathcal{R},E) \mathrm{d}\mathcal{R}$, where $N_{\rm TD,b}(E)$ is the number of bound stars that eventually enter the tidal radius
with energy between $E$ and $E+\mathrm{d}E$.
The quantity $\mathcal{R}$ and $E$ in equation~\ref{Eq-n_TD(R,E)} shall take the values at the disruption, because these values are relevant to our theoretical models of the $e$ and $\beta$ distributions. The upper cutoff of $\mathcal{R}$ comes from the condition for tidal disruption: the separation between a star and the SMBH must be less than or equal to $r_{\rm t}$. This condition is translated to $\mathcal{R}\leq\mathcal{R}_{\rm lc}$ at the disruption according to the equations~\ref{eq:angmom},~\ref{Eq:r_p-J2} and~\ref{Eq:r_t-Jlc2}.
The value of $N_{\rm TD,b}(E)$ results from the
accumulation of TDEs with time, thus it is calculated as $N_{\rm TD,b}(E)=\int F(E,t)\mathrm{d}t$, where $F(E,t)$ is the flux of stars that enter the loss-cone at time $t$. In this work $N_{\rm TD,b}(E)$ is obtained directly from the $N$-body simulation (for example using the orbital energy of the TDEs recorded in the simulation, see Figure~\ref{fig_NTD,b}),
so we do not discuss $F(E,t)$ here in detail---but see e.g. Section 2.2 in \cite{Merritt2013}. Note that the normalization of all number density distributions of tidally disrupted stars presented in this paper is to the total number of disrupted stars over the time of the simulation.
From $N_{\rm TD,b}(E)$ we obtain for the normalization coefficient
\begin{equation}
A_{\rm TD,b}(E) =
\frac{N_{\rm TD,b}(E)}{
\mathcal{R}_{\rm lc}\ln\mathcal{R}_{\rm lc}
-\mathcal{R}_{\rm lc}\ln\mathcal{R}_{\rm 0}
-\mathcal{R}_{\rm lc}+\mathcal{R}_{\rm 0}
}.
\label{Eq-A_TD(E)}
\end{equation}
By introducing
\begin{equation}
Q \equiv \frac{\Delta\mathcal{R}}{\mathcal{R}_{\rm lc}},
\label{eq:q}
\end{equation}
where $\Delta\mathcal{R}$ is the cumulative change of $\mathcal{R}$ over one orbital period of the star,
\cite{MT1999} evaluated $\mathcal{R}_{\rm 0}$ by
\begin{equation}
\mathcal{R}_{\rm 0} = g(Q) \mathcal{R}_{\rm lc},
\label{Eq-R0}
\end{equation}
where 
\begin{equation}
g(Q)=\mathrm{exp}\left[-(Q^4+Q^2)^{1/4}\right]
\label{Eq-g(Q)}
\end{equation}
is an approximation of the analytical solution derived by \cite{Merritt2013}, whose exact form is expressed in terms of Bessel series [also see equations (44,45) in \cite{VM2013}].
Since $\mathcal{R}_{\rm 0}$ is very close to $\mathcal{R}_{\rm lc}$ in the case of $Q<1$,
the number density almost goes to zero in the loss-cone region. $Q<1$ is the empty loss-cone regime; 
while in the $Q>1$ case $\mathcal{R}_0\ll{R}_{\rm lc}$, we are in the full loss-cone regime. We will explain how to compute $Q$ in Section~\ref{SECT-TEST}.

\subsection{Number density of bound stars}
\label{Theory-bound-star}
In this subsection we transform the number density of bound stars in the loss-cone according to equation \ref{Eq-n_TD(R,E)} from the standard phase space variables $\mathcal{R}$ and $E$ into new variables more suitable for our analysis of TDEs. 
We consider transformations into the following new pairs of
variables: $(\mathcal{R},E)$ into
$(e,E)$, $(\beta,E)$ or $(e,\beta)$. 
All Jacobian determinants are nonsingular (see Table~\ref{table-Jacob}), so all pairs can be used as new independent phase space variables. We focus in the following on
$(e,E)$ and $(\beta,E)$---after integration over $E$ the resulting distributions in $e$ and $\beta$ can be compared with $N$-body simulations and also be used to analyze expected TDE characteristics (see Section~\ref{SECT-TEST}).
The details of the variable transformation are presented in Appendix~\ref{Appendix-relation}, which are derived in the Keplerian regime. For the marginally bound stars ($E\approx 0$), the Keplerian assumption breaks down and the variable transformations given in the Appendix may become inaccurate.

Substituting the variable $\mathcal{R}$ in the expression of $n_{\rm TD,b}(\mathcal{R},E)$
(equation~\ref{Eq-n_TD(R,E)}) with $1-e^2$ (equation \ref{Eq-R-e})
and multiplying the corresponding Jacobian determinant provides
the number density of the bound stars in the range of $e_{\rm ll} \leq e \leq e_{\rm ul}$:

\begin{equation}
N_{\rm TD,b}(e,E)
= 2 \, A_{\rm TD,b}(E)\, e \, \ln\left[\frac{(1\!-\!e^2)}{g(Q)}\cdot \frac{|E_{\rm t}|}{2|E|}\right],
\label{Eq-n(e,E)-bound}
\end{equation}
where
\begin{equation}
e_{\rm ll}= \sqrt{1 \! - \! 2 \, \frac{|E|}{|E_{\rm t}|}} \ \ ; \ \ 
e_{\rm ul}= \sqrt{1 \! - \! 2 \, g(Q)\frac{|E|}{|E_{\rm t}|} }
\label{eq:ell}
\end{equation}
are the boundaries of eccentricity obtained from the $\mathcal{R} = \mathcal{R}_{\rm lc}$ limit and
$\mathcal{R} = \mathcal{R}_{\rm 0}$.
Note that $n_{\rm TD,b}(e,E)$ goes to zero outside of this eccentricity range.

Substituting the variable $\mathcal{R}$ in the expression of
$N_{\rm TD,b}(\mathcal{R},E)$ with
$2|E|/(|E_{\rm t}|\beta)$ (equation \ref{Eq-beta})
and multiplying the corresponding Jacobian determinant, we obtain
\begin{equation}
N_{\rm TD,b}(\beta,E) = \frac{A_{\rm TD,b}(E)}{\beta^2}\frac{2|E|}{|E_{\rm t}|}
[-\ln g(Q) - \ln\beta]
\label{Eq-n(beta,E)-bound}
\end{equation}
in the range of $1\leq \beta \leq 1/g(Q)$, where we used equations (\ref{Eq-n_TD(R,E)})
and (\ref{Eq-R0}) for the derivation. Note that the number density vanishes to zero
at $\beta = 1/g(Q)$, which corresponds to that $\mathcal{R}$ equals $\mathcal{R}_{\rm 0}$
in the original number density (see equation~\ref{Eq-n_TD(R,E)}).
For $Q\gg1$, if $\ln\beta$ is negligible compared to $-\ln g(Q)$,
equation~\ref{Eq-n(beta,E)-bound} can be approximated as
$N_{\rm TD,b}(\beta,E) \simeq 2Q|E|A_{\rm TD,b}(E)/(|E_{\rm t}|\beta^2)$, verifying the $\beta^{-2}$ dependence in the full loss-cone regime suggested by~\cite{SM2016}.

We also notice that for $Q=1$ ($g(Q) = 1/{\rm e}$; e: Euler's number), which corresponds to an energy value at the critical radius \citep{FR1976, 2004MNRAS.352..655A}, for fixed critical energy $E=E_{\rm crit}$ we get the following results:
\begin{equation}
N_{\rm TD,b}(\beta,E_{\rm crit}) \propto 
\beta^{-2} \cdot \left( 1 \!-\! \ln\beta\right) \ ; \ 
N_{\rm TD,b}(e,E_{\rm crit})  \propto 
e \cdot \ln\left(1\!-\!e^2\right) \ . 
\end{equation}

%
\subsection{Number density of unbound stars}
\label{Theory-unbound-star}
After the pioneering work by \cite{CK1978,SM1978} for bound
stars there were more general papers, extending the domain of solution of the Fokker-Planck equation to the unbound stars in the galaxy. They focused
on the tidal disruption event rate and studied the dependence of the event rate on the
geometry of the host cluster~\citep{MT1999}, the $M_{\rm BH}-\sigma$ relation~\citep{WM2004} and the stellar
mass spectrum~\citep{SM2016}. All of them used the standard phase space variables of energy and angular momentum.
The stellar number density in that case is written as
$n(E_{\rm tot},\mathcal{R})$, with $E_{\rm tot} = v^2/2 - GM_{\rm BH}/r+ \Phi_{\rm gal}(r)$; the new term $\Phi_{\rm gal}(r)$ denotes the gravitational potential generated by all stars of the nuclear star cluster and the galaxy inside a radius $r$ (here, for example, in the case of a spherical system).
In the following we will argue that a direct variable transformation to our variables $e$ and $\beta$ as before is cumbersome and actually not really necessary. Let us first check the function $\mathcal{R}(\beta,E_{\rm tot})$, which defines the variable transformation from $\mathcal{R}$ and $E_{\rm tot}$ to $\beta$ and $E_{\rm tot}$.
For $\beta$ we need to find the pericenter distance $r_{\rm p}$ as a function of $E_{\rm tot}$ and $\mathcal{R}$.
From our definition of  $\mathcal{R}$ in equation~\ref{eq:angmom} we get from the expression for $E_{\rm tot}$ above:
\begin{equation}
\mathcal{R} \frac{J^2_{\rm c}(E_{\rm tot})}{2r^2} - \frac{GM_{\rm BH}}{r} + \Phi_{\rm gal}(r) - E_{\rm tot} = 0.
\end{equation}
$r_{\rm p}$ is the smallest root of this equation in terms of $r$. Since that depends on the functional form of 
$\Phi_{\rm gal}(r)$ it is generally impossible to find an analytic solution; it has to be computed numerically for each galaxy.

In the case of eccentricity, another problem occurs---the definition of $e$ for the two-body problem has no straightforward generalization for orbits in a more general star cluster or galactic potential. Typically, orbits in galactic potentials are not closed; generalized eccentricities may be defined using the angular momentum or pericenter and apocenter ($r_{\rm a}$), $e = (r_{\rm a}-r_{\rm p})/(r_{\rm a}+r_{\rm p})$, but it is not always a conserved quantity except for in a spherical potential. For stars, which we are interested in, when they come close to the SMBH, the two-body eccentricity will be different from a value computed far out in the galaxy.
Therefore, we look at the situation of a two-body problem only, for a hyperbolic encounter between a star and the SMBH.
We compute $e$ at a place near the tidal radius, and
convert the orbital energy of the star to $E=E_{\rm tot}-\Phi_{\rm gal}(r_{\rm t})$, which is positive for an unbound star.
Adopting the relation between $e$ and $\beta$ for the hyperbolic orbit (equation~\ref{Eq-beta-e-unbound}), we obtain

\begin{equation}
e
=1+\frac{E_{\rm tot}-\Phi_{\rm gal}(r_{\rm t})}{|E_{\rm t}|}\frac{1}{\beta(\mathcal{R},E_{\rm tot})}.
\end{equation}
\noindent
There is no simple and universal relation between $e$, $E_{\rm tot}$ and $\mathcal{R}$, due to the complicated $\beta(\mathcal{R},E_{\rm tot})$ term, resulting in a complex expression also for the
transformation $\mathcal{R}(e,E_{\rm tot})$. Therefore, we do not use the number density of unbound stars in the form of \cite{MT1999}, \cite{WM2004} or \cite{SM2016}.

Instead, we use a simpler approximation for the number density of unbound stars in terms of $e$ and $\beta$ near the tidal radius,
which is also appropriate for the comparison with our $N$-body results (see Section~\ref{SECT-TEST}).

Outside of the SMBH influence radius $r_{\rm h}$ the loss-cone has negligible effect on the stellar distribution, because it is usually $r_{\rm t}\ll{r}_{\rm h}$. 
Therefore the velocity distribution is close to a Gaussian along the principal axes of a velocity ellipsoid, also denoted as anisotropic Schwarzschild distribution; it allows for different velocity dispersions, e.g., in radial and tangential directions---for star clusters, see \cite{2004MNRAS.352..655A}, but see also \cite{Kazantzidisetal2004} for a counterexample in the galactic nuclei.
In the following, we derive the $N(\beta)$ based on a simple cross-sectional ansatz. The cross section for the stars that could pass within a distance $r$ from the SMBH (with gravitational focusing) is $\Sigma(r)=\pi r^2 [1+(GM_{\rm BH}/r)/(v^2/2)]$. When the star's specific kinetic energy $v^2/2$ at the orbital apocenter is much smaller than the gravitational potential $GM_{\rm BH}/r$ at the pericenter of its orbit, as is the case discussed here, the cross section is approximately reduced to be $\Sigma(r)\approx 2\pi r GM_{\rm BH}/v^2$. Then the flow of stars that passes within $r_{\rm p}$ can be estimated as $f(<r_{\rm p})=n \Sigma v \approx 2\pi r_{\rm p} n GM_{\rm BH}/v$, where $n$ and $v$ are the stellar number density and stellar velocity at the place from where these stars come. Thus we get the relation $f(<r_{\rm p})\propto(r_{\rm p}/r_{\rm t})r_{\rm t}$ (see also equation 2 of \cite{R1988}, but note that we do not need to postulate isotropy here - it is sufficient to use the radial velocity only, because the tangential velocity is very small for loss-cone stars originating from a large distance to the black hole. Substituting $\beta=r_{\rm t}/r_{\rm p}$ into the above relation, we find $f(>\beta)\propto\beta^{-1}$. By definition, $f(>\beta)$ is computed as $\int_{\beta}^{\infty} N(\beta)\mathrm{d}\beta$, hence the number density $N(\beta)$ is proportional to $\beta^{-2}$ and we write down the following expression:
\begin{equation}
N_{\rm TD,u}(\beta,E) = A_{\rm TD,u}(E)\beta^{-2},
\label{Eq-n(beta,E)-unbound}
\end{equation}
in the range of $\beta\geq1$.
Equating $\int_1^{\infty} N_{\rm TD,u}(\beta,E) \mathrm{d}\beta$ to
$N_{\rm TD,u}(E)$ results in the normalization coefficient
$A_{\rm TD,u}(E) = N_{\rm TD,u}(E)$.

Then substituting the variable $\beta$ in the expression of $N_{\rm TD,u}(\beta,E)$ with $E/[|E_{\rm t}|(e-1)]$ (equation \ref{Eq-beta-e-unbound}) and multiplying
the corresponding Jacobian determinant, we have

\begin{equation}
N_{\rm TD,u}(e,E)
=A_{\rm TD,u}(E)\frac{|E_{\rm t}|}{E}
\label{Eq-n(e,E)-unbound}
\end{equation}
in the range of $1 < e \leq 1+ E/|E_{\rm t}|$.
Note that $N_{\rm TD,u}(e,E)=0$ if $e>1+E/|E_{\rm t}|$.
%

%
\section{Comparison with numerical experiments}
\label{SECT-TEST}
%
Here we compare the distribution of tidally accreted stars in terms of $e$ and $\beta$, which we have analytically derived in the preceding section, with $N$-body simulations.
To get better statistical quality of the results we use in these comparisons the dependence on $e$ and $\beta$ only, rather than the joint 2D distribution in $e$, $\beta$ (or a 3D distribution in $E$, $e$, $\beta$), for the reason of statistical noise, by integrating over all energies as follows:
\begin{equation}
N_{\rm TD,b}(e) = \int_{E_{\rm t}}^{0} N_{\rm TD,b}(e,E) \mathrm{d}E \ \ ; \ \ 
N_{\rm TD,b}(\beta) = \int_{E_{\rm t}}^{0} N_{\rm TD,b}(\beta,E) \mathrm{d}E \ ,
\label{Eq-n(beta)-bound}
\end{equation}
where we use the number densities of equations~(\ref{Eq-n(e,E)-bound}) and (\ref{Eq-n(beta,E)-bound}).
For the number densities of
unbound stars, we obtain $N_{\rm TD,u}(e)$ and $N_{\rm TD,u}(\beta)$ in the same way, but integrate from $E=0$ to infinity.
To evaluate the number densities from these equations we need to get an evaluation for $Q$. Equation~\ref{eq:q} shows that it requires the computation of the average angular momentum change per orbit $\langle\Delta J^2\rangle $.
To measure $\langle\Delta J^2\rangle $ from the simulation, one needs to record positions and velocities of all particles at very high frequency. We did not save such data from our models, but note that \cite{VM2013} have done such measurements and the results generally agree with the theoretical prediction, but have large scatters (see their Figure 8). Hence, we turn to the analytical solution of $\langle\Delta J^2\rangle $ to construct our theoretical model.

$\langle\Delta J^2\rangle $ is computed at the apocenter of a stellar orbit, because two-body relaxation affects the orbit most strongly at the apocenter passage.
This assumption can be justified because the orbiting star passes its apocenter so slowly that it has more time to interact with the surrounding stars, and also the perturbing forces
may exert a non-negligible torque to the passing star \citep{TT1997,ZBS2015}.
From relaxation theory \citep{FR1976,Merritt2013}, we get
\begin{equation}
\langle\Delta J^2\rangle \approx J_{\rm c}(r_{\rm a})^2\left(\frac{t_{\rm dyn}}{t_{\rm relax}}\right),
\label{eq:meansqj}
\end{equation}
where $J_{\rm c}(r_a)=\sqrt{GMr_a}$ is the specific angular momentum of a circular orbit at $r_a$,
$t_{\rm dyn}=r_a / \sigma(r_a)$
is the dynamical timescale,
\begin{equation}
t_{\rm relax}=\frac{0.34\sigma^3(r_{\rm a})}{G^2 m \rho(r_{\rm a})\ln\Lambda}.
\label{Eq-t_relax}
\end{equation}
is the local relaxation time \citep{Spitzer1987}, $\sigma(r)$ is the velocity dispersion of
the stars, $\rho(r)$  is the radial density profile of the star cluster, $m=M_{\rm c}/N$ is the
mass of the star, and $\ln\Lambda = \ln(0.11N)$ is the Coulomb logarithm~\citep{GS1994}.
Note that equation~\ref{eq:meansqj} is correct only qualitatively, as there are some cases where the assumption used in equation~\ref{eq:meansqj} is invalid; e.g., in the ultrasteep stellar cusps~\citep{FS2018,SGV2018}, but such cusps are not presented in our simulations.
We introduce a dimensionless parameter, $Q_{\rm boost}$, of order unity to provide an approximate evaluation of $Q$. From equations (\ref{eq:q}) and (\ref{eq:meansqj}), $Q$ is then given by

\begin{equation}
Q=\frac{\Delta{R}}{R_{\rm lc}}=Q_{\rm boost}\left(\frac{t_{\rm dyn}}{t_{\rm relax}}\right)\left(\frac{r_a}{2r_{\rm t}}\right).
\label{eq:appq}
\end{equation}
Here we have used $\mathcal{R}_{\rm lc}=J_{\rm lc}^2/J_c(r_a)^2$. One can approximate $r_a=a(1+e)\approx 2a$ for bound stars on highly eccentric orbits in the Keplerian potential, with the orbit's semimajor axis $a$. Since $a = -GM_{\rm BH}/(2E)$, we conclude that
for the given density profile and velocity dispersion, $Q$ and $g(Q)$ become a function of only
$E$ because of equation~(\ref{Eq-g(Q)}).
However, the quantity $-GM_{\rm BH}/(2E)$ diverges as $E$ approaches 0. In practice, we compute the exact value of $r_a$ from the combined gravitational potential $\phi(r)$.

Note that for unbound tidally accreted stars, according to equations~\ref{Eq-n(beta,E)-unbound} and \ref{Eq-n(e,E)-unbound}, the number density does not depend on $Q$.
%

%
\subsection{Basic model}
\label{SECT-Basic_model}

%
For comparison of $N$-body simulations with the analytical results, we use the data of our previously published study \citep{HZL2018}. We choose from that paper two models; each has particle number $N = 512$K and $r_{\rm t}=10^{-5} $. They are the models with largest particle number and smallest tidal radius in that parameter study, we consider them as the ones closest to a real nuclear star cluster (though still not sufficient in terms of particle number). The two models differ only by their black hole mass: one has $M_{\rm BH}=0.01$, while the other has $M_{\rm BH}=0.05$ (referring to models 5 and 10 of \citet{HZL2018}, respectively), in units where the total cluster mass is unity.

Our spherical star cluster with $N$ equal-mass stars and a star-accreting SMBH fixed at the center was initialized in the same way as in our previous papers \citep{HZL2018,ZBS2014}---initially a Plummer model was used, which has a central flat core, which adjusts to
the gravity of the central back hole during a few dynamical orbits, producing a cusp-like initial density distribution. More details about the time evolution and the profiles of density and velocity dispersion can be found in \cite{ZBS2014}. We use dimensionless H\'{e}non units, in which $G=M_{\rm c}=1$ and the
total energy of the system is $E = -1/4$~\citep{Heggie2014b,Heggie2014a}. In the
simulations, we take $r_{\rm t}$ as a fixed accretion radius, in which all the stars are
regarded as being tidally disrupted and removed from the simulations. More details can also be found in~\citet{HZL2018}.
Our $N$-body models adopt initially isotropic velocity distribution, therefore some stars are placed inside the loss-cone at the beginning. These stars shall cause a burst of TDEs. However, such a burst cannot last for a long time, because the loss-cone runs out of stars within one orbital period (at most, a few $N$-body time units), then the system enters the angular-momentum-diffusion-dominated phase. As a result, such initial surge of TDEs only accounts for less than a few percent of all the TDEs, hence, their impact on the validation of our theoretical models are negligible. 
After the initial adjustment, a central density cusp is established in the $N$-body simulation, even though the total simulation has been only about one-third of a half-mass relaxation time. 
Although the simulation times of these two models were less than one-third of the half-mass relaxation time, \cite{PMS2004} have shown that this is enough for the system to achieve the CK78 distribution.

Before approaching our final goal, to compare the number densities according to equations~\ref{Eq-n(beta)-bound} with $N$-body simulations, we will first check the quantities $Q=Q(E)$, $A_{\rm TD,b}(E)$ and $A_{\rm TD,u}(E)$ because they are required for
the calculation of the analytical number densities. According to
the definition, $A_{\rm TD,b}(E)$ depends on $N_{\rm TD,b}(E)$ and $Q$ (equation~\ref{Eq-A_TD(E)})
, while $A_{\rm TD,u}(E)$ just equals to $N_{\rm TD,u}(E)$.
In order to evaluate these quantities, we measure
$N_{\rm TD,b}(E)$ and $N_{\rm TD,u}(E)$ and approximate it by double-power-law function. Figure~\ref{fig_NTD,b} illustrates the results obtained from the $N$-body model with $M_{\rm BH} = 0.01$, measured at the end of the simulation.
In both panels, the red histograms shows $N$-body data of bound and unbound tidally disrupted stars, as a function of their energy. The stars are distributed between $E_{\rm min}$ ($<0$) and $E_{\rm max}$ ($>0$).
Note that $E_{\rm min}$ (roughly $-2$ in model unit) is much larger than $E_{\rm t}$ ($-500$ in model unit). This is consistent with the loss-cone theory that the stars are originating far from the tidal disruption radius. $N_{\rm TD,b}(E)$ and $N_{\rm TD,u}(E)$ is used to compute the analytical expressions for $A_{\rm TD,b}(E)$ and $A_{\rm TD,u}(E)$.

To get an analytic expression for $Q$ we use a similar method as before to approximate now the density profile and velocity dispersion profile of the star cluster using $N$-body data. To model the density profile, we use the following double-power-law function to fit the $N$-body data
\begin{equation}
\rho(r)=\rho_0\left(\frac{r}{r_0}\right)^{\gamma}
\left[1+\left(\frac{r}{r_0}\right)^{\alpha}\right]^{\frac{\gamma-\delta}{\alpha}}.
\label{eq:den_profile}
\end{equation}
Then the 1D velocity dispersion is obtained via the Jeans equation (with $G=1$),
\begin{equation}
\rho(r)\sigma^2(r)=\int_r^{\infty} [M_{\rm BH}+ M_{\star}(<x)]\rho(x)x^{-2} \mathrm{d}x.
\label{eq:sig2_by_JeansEq}
\end{equation}

An example of the density and velocity dispersion profiles is depicted in
Figure~\ref{fig_sig-rho}, which are measured from the simulation data when the density profile is stabilized. We also show the results of the double power law fitting on the density profile and the solution of Jeans equation for the velocity dispersion, which smooth the fluctuations in the data and are used to calculate $Q$. We observe that in the central part, well inside the influence radius ($r\leq 0.03$), the simulated density profile show a steeper cusp (although very noisy due to low particle number) than the prediction of double-power-law fitting. This deviation only mildly affects our modeling, since the stellar mass in this cusp is less than $10^{-3}$ and almost none of the disrupted stars are originated from this region.
In our $N$-body model, the influence
radius is defined as the radius within which the enclosed stellar mass equals the SMBH mass.
The influence radius in the model with $M_{\rm BH}=0.01$ ($M_{\rm BH}=0.05$)
roughly equals 0.1 (0.2).

From the fitted density profile (equation~\ref{eq:den_profile}), we also compute the composited gravitational potential in the star cluster $\phi(r)$ and the apocenter $r_a$ for the (zero angular momentum) radial orbit with a given orbital energy $E$. An example of $r_a(E)$ is shown in the right panel of Figure~\ref{fig_sig-rho}.

Figure~\ref{fig_Q-r_a} depicts $Q$ as a function of $r_a$, which is evaluated by equation~\ref{eq:appq}. Here, the density and velocity dispersion profile are modeled by equations~\ref{eq:den_profile} and \ref{eq:sig2_by_JeansEq} (see also Figure~\ref{fig_sig-rho}). For comparison purposes, we adopt two different values of $Q_{\rm boost}=1$ and $Q_{\rm boost}=5$. From the figure, we find that both the empty loss-cone regime ($Q<1$) and full loss-cone regime ($Q>1$) are present in our $N$-body data. The above criteria for the empty and full loss-cone regime are obtained by comparing the size of the loss-cone and the size of angular momentum diffusion (see equation~\ref{eq:q}). \citet{MT1999} have proposed another criterion, where the loss-cone regimes are separated by $Q\simeq-\ln\mathcal{R}_{\rm lc}$, which comes from the consideration of the loss-cone flux. This alternative criterion does not change our conclusion that both the empty and full loss-cone regimes exist in our $N$-body models.

\begin{figure}[htbp]
  \begin{center}
  \includegraphics[width=0.47\columnwidth]{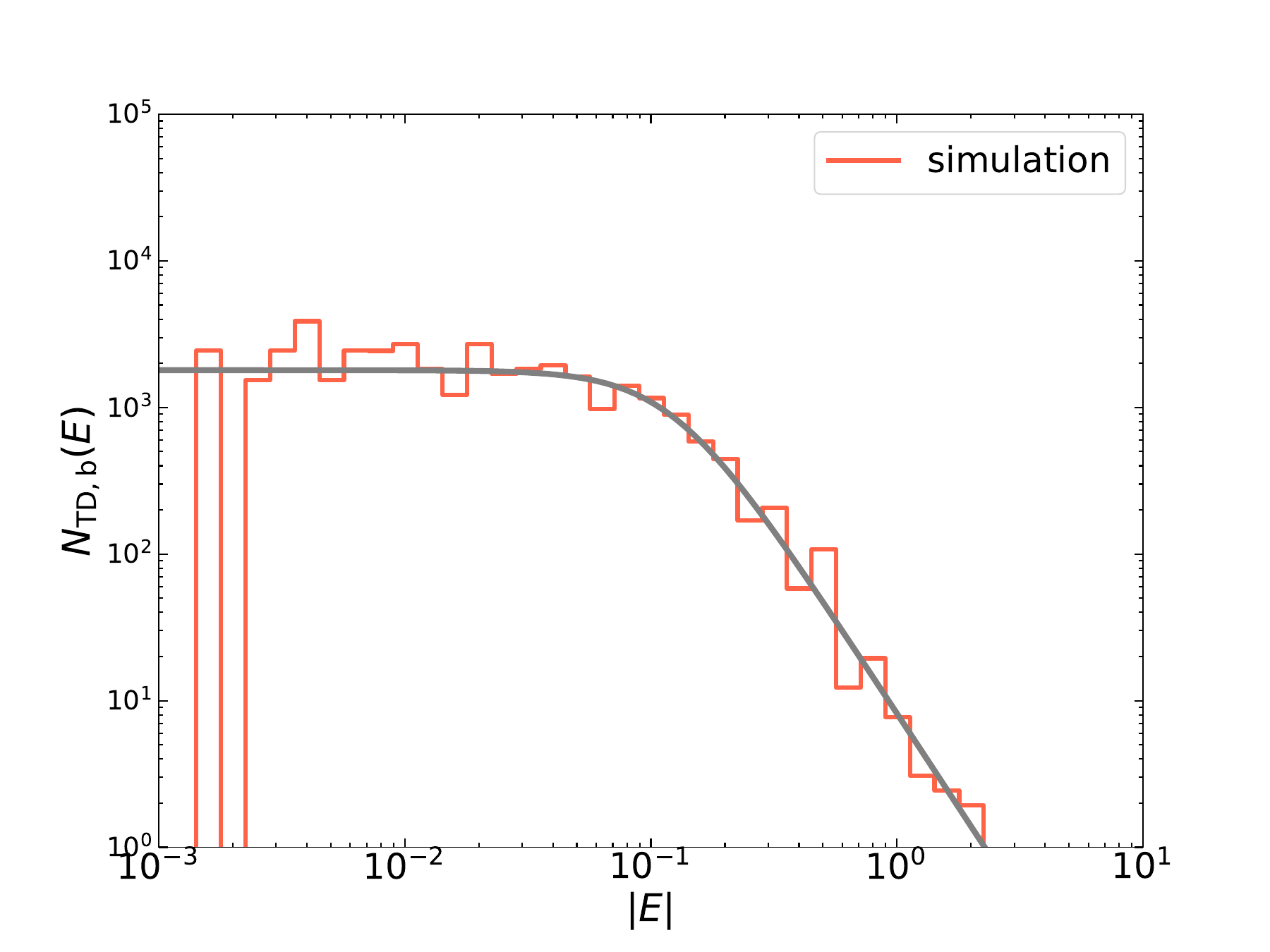}
  \includegraphics[width=0.47\columnwidth]{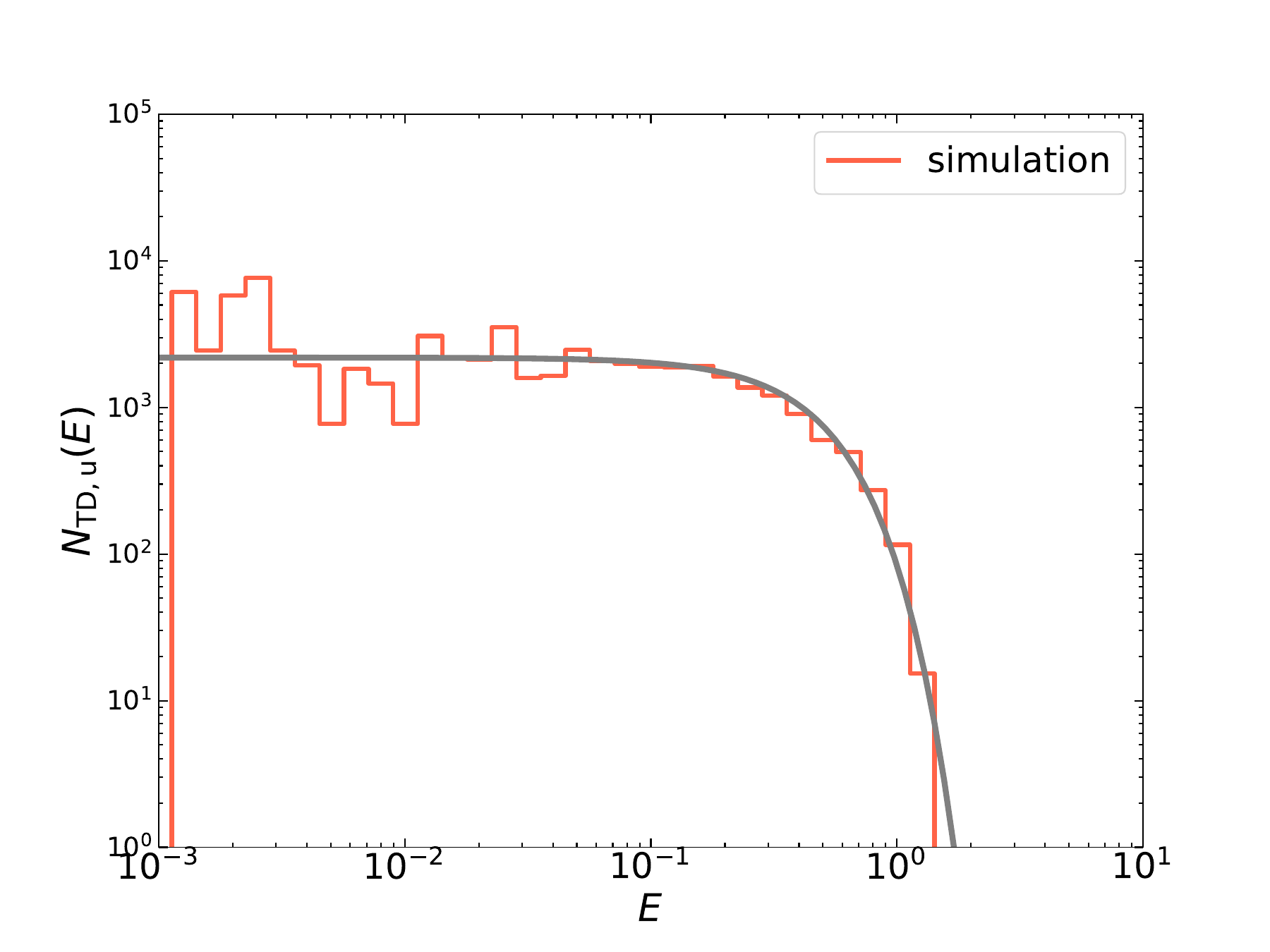}
  \end{center}
 \caption{
Energy dependence of the cumulative number densities for both the bound (left panel)
and unbound (right panel) disrupted stars with $M_{\rm BH} = 0.01$ cases. In each panel,
the red histogram shows the simulated number density ($\mathrm{d}N/\mathrm{d}E$), whereas the gray line represents the double-power-law curve, which is fitted to the simulated number density using equation taking the same functional form of equation (\ref{eq:den_profile}), but replacing the density and radius quantities with number density and energy, respectively.
}
\label{fig_NTD,b}
\end{figure}

\begin{figure}[htbp]
  \begin{center}
  \includegraphics[width=0.32\columnwidth]{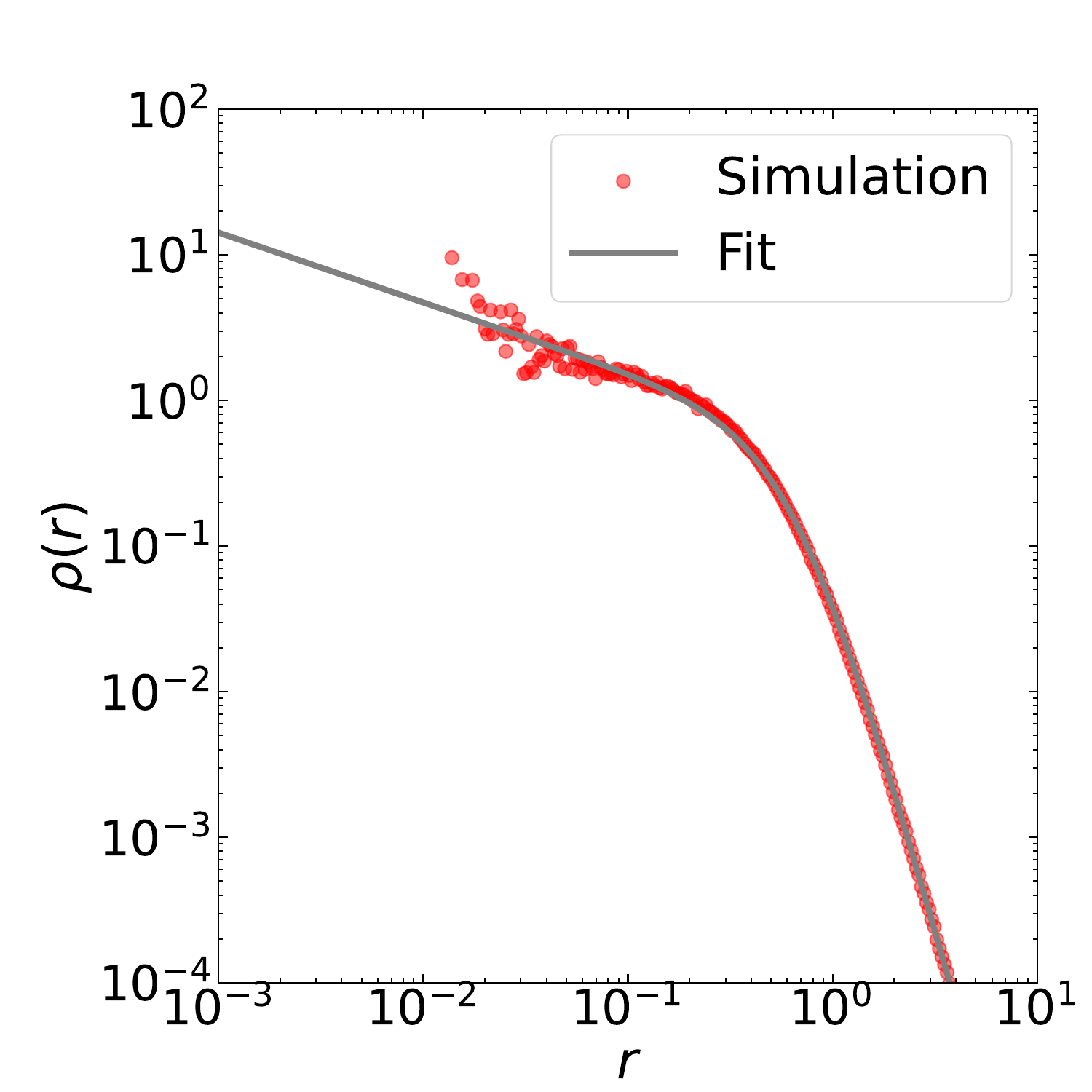}
  \includegraphics[width=0.32\columnwidth]{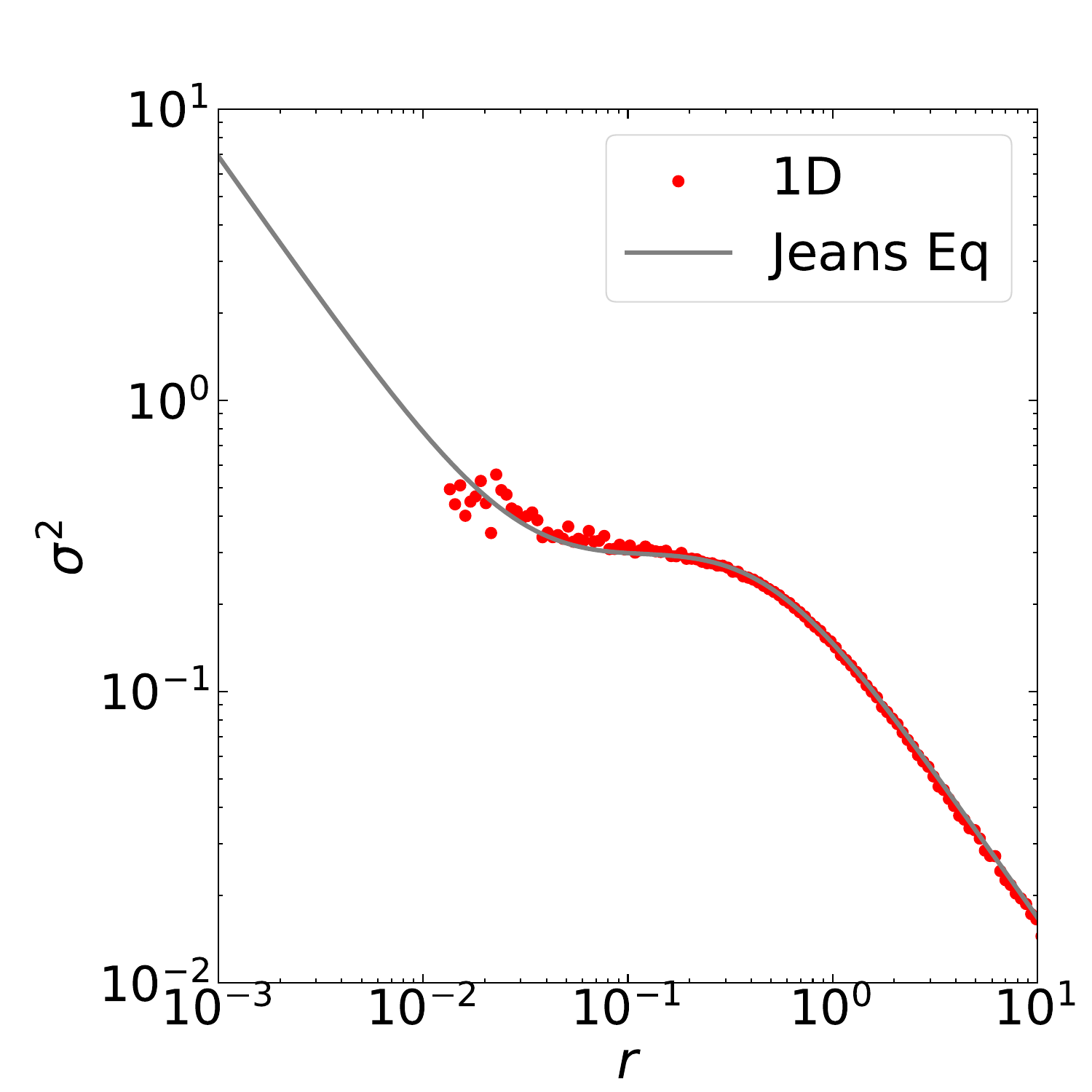}
  \includegraphics[width=0.32\columnwidth]{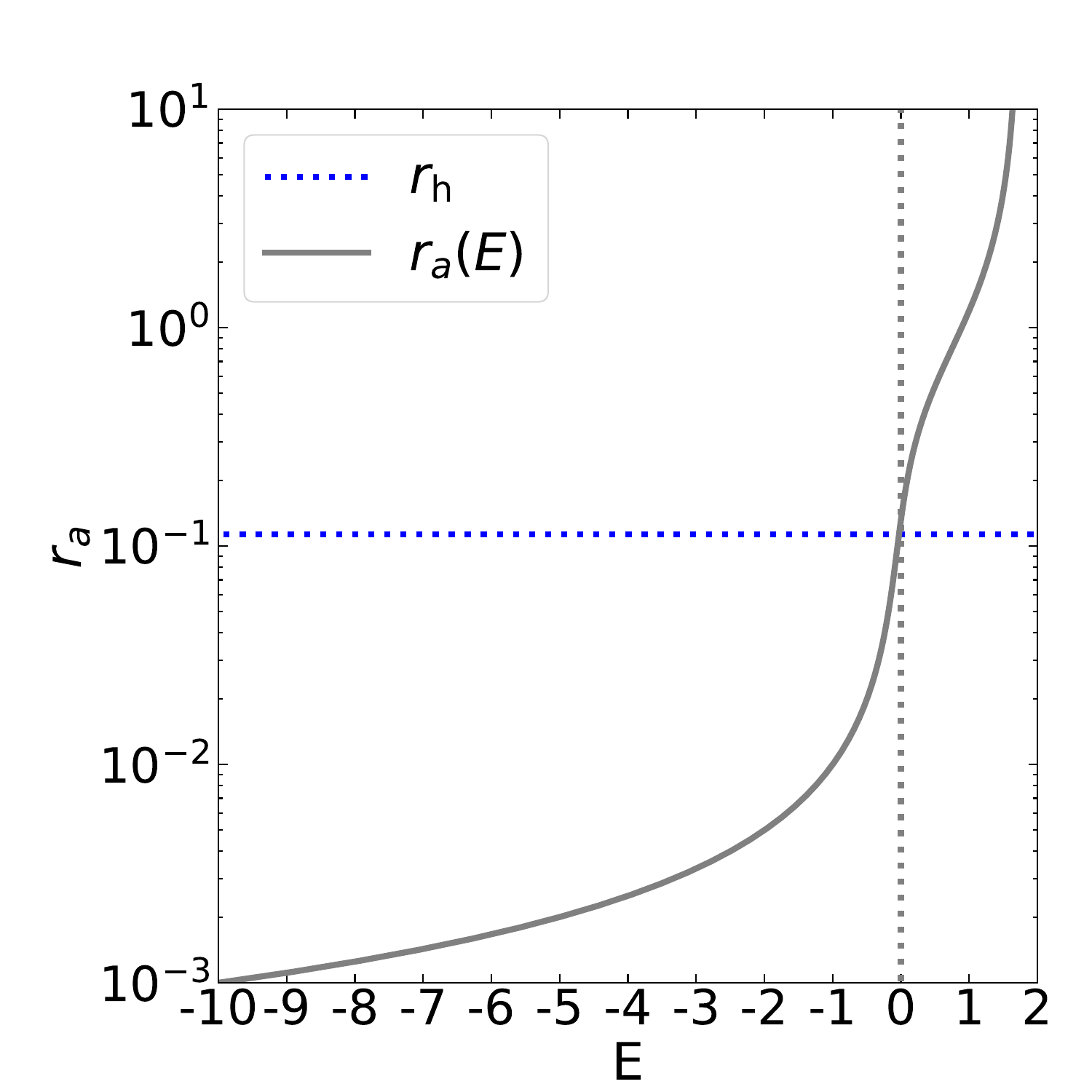}
  \end{center}
  \caption{
Radial profiles of the density (left) and of the square of the stellar velocity dispersion (middle) for the model cluster with $M_{\rm BH}=0.01$. In the left panel, the red dots and gray solid line represent the simulated and numerically fitted density profiles, respectively. In the middle panel, the red dots and gray solid line denote the square of stellar velocity dispersions evaluated by the simulation and by equation~(\ref{eq:sig2_by_JeansEq}), i.e., the Jeans equation, respectively. The right panel depicts the apocenter distance radius $r_a$ as a function of the specific orbital energy $E~(=v^2/2-GM_{\rm BH}/r)$ for a radial orbit with zero angular momentum. The horizontal blue dotted line indicates the influence radius $r_{\rm h}$, whereas the vertical black dotted line corresponds to the line of $E=0$.
}
\label{fig_sig-rho}
\end{figure}

\begin{figure}[htbp]
  \begin{center}
  \includegraphics[width=0.8\columnwidth]{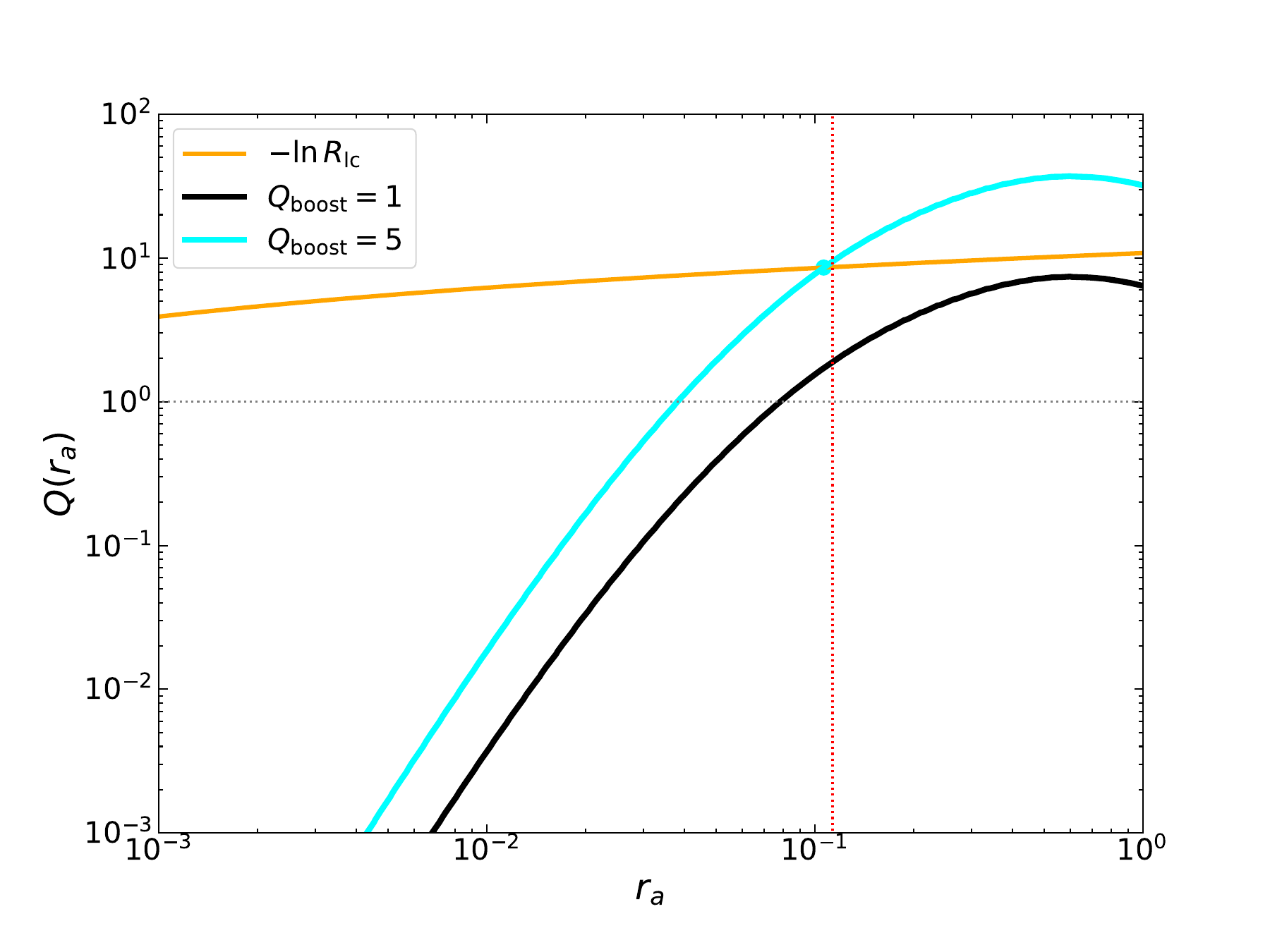}
  \end{center}
  \caption{
  Dependence of $Q$ on the apocenter distance radius $r_{\rm a}$ for bound stars. H\'{e}non units are adopted for the plot. The solid ocher line and horizontal black dotted line delineate the $-\ln\mathcal{R_{\rm lc}}$ curve ( see \citealt{MT1999} for details) and the $Q=1$ line, respectively, which provides the criterion for distinguishing the empty and full loss-cone regimes. The vertical dotted red line represents the influence radius of the SMBH. The solid black and light blue lines indicate the $Q$ curves with $Q_{\rm boost}=1$ and $Q_{\rm boost}=5$, respectively, which are given by applying the fitted simulation results to equation~(\ref{eq:appq}).
}
\label{fig_Q-r_a}
\end{figure}

%
\subsection{Distribution of tidally accreted stars in eccentricity and penetration factor}
%

\begin{figure}[htbp]
  \begin{center}
  \includegraphics[width=0.47\columnwidth]{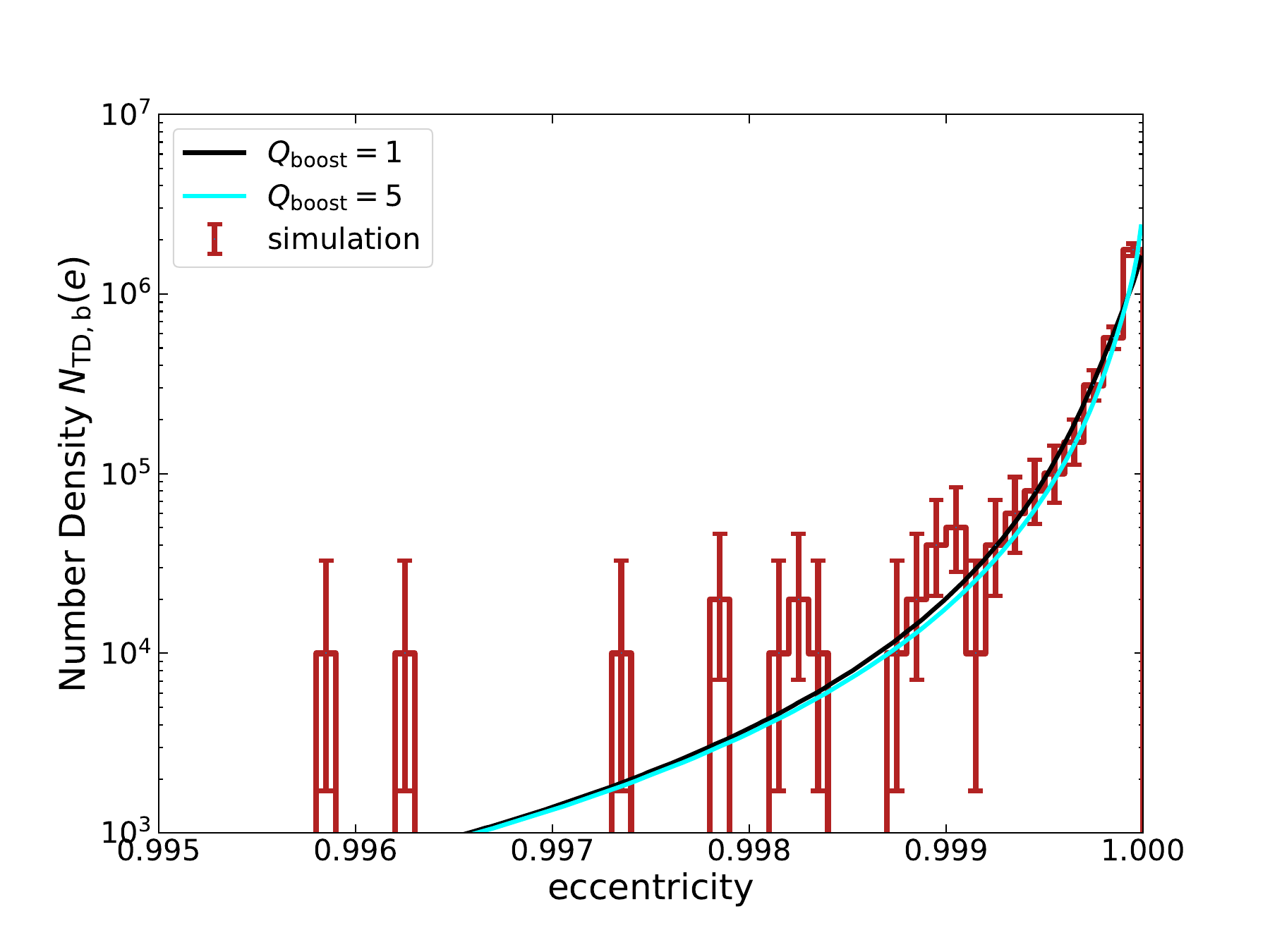}
  \includegraphics[width=0.47\columnwidth]{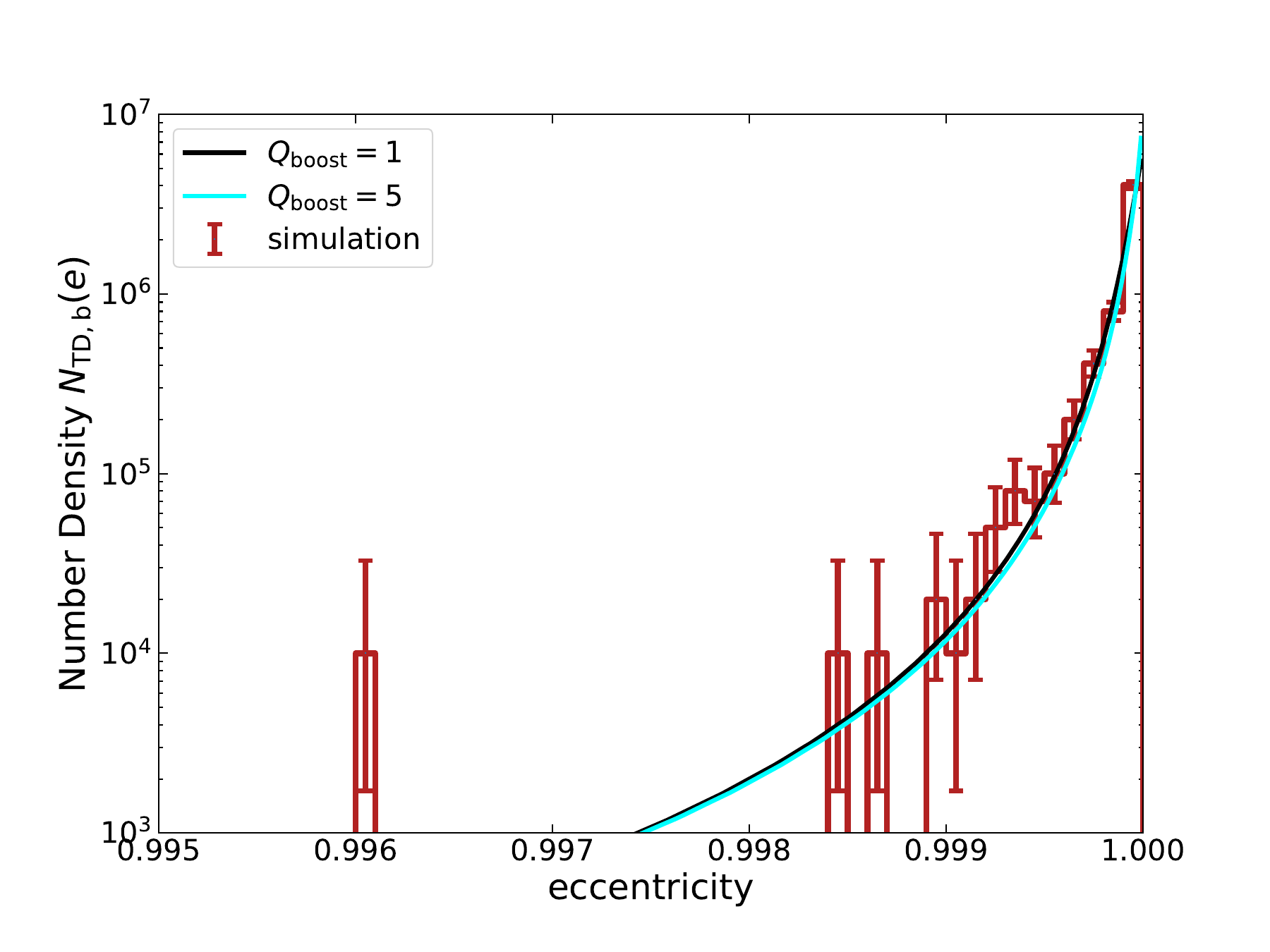}
  \end{center}
 \caption{
Dependence of the number density of the bound stars on the orbital eccentricity. The red histogram represents the simulated number density, while the solid black and light blue lines delineate the theoretical number densities with $Q_{\rm boost}=1$ and with $Q_{\rm boost}=5$, respectively. Note that the theoretical number density is computed by equation~(\ref{Eq-n(beta)-bound}). The left panel corresponds to the $M_{\rm BH} = 0.01$ case, while the right panel corresponds to the $M_{\rm BH} = 0.05$ case. The error bars indicate the statistical uncertainty corresponding to the standard deviation.
}
\label{fig_n_e_b}
\end{figure}

\begin{figure}[htbp]
  \begin{center}
  \includegraphics[width=0.47\columnwidth]{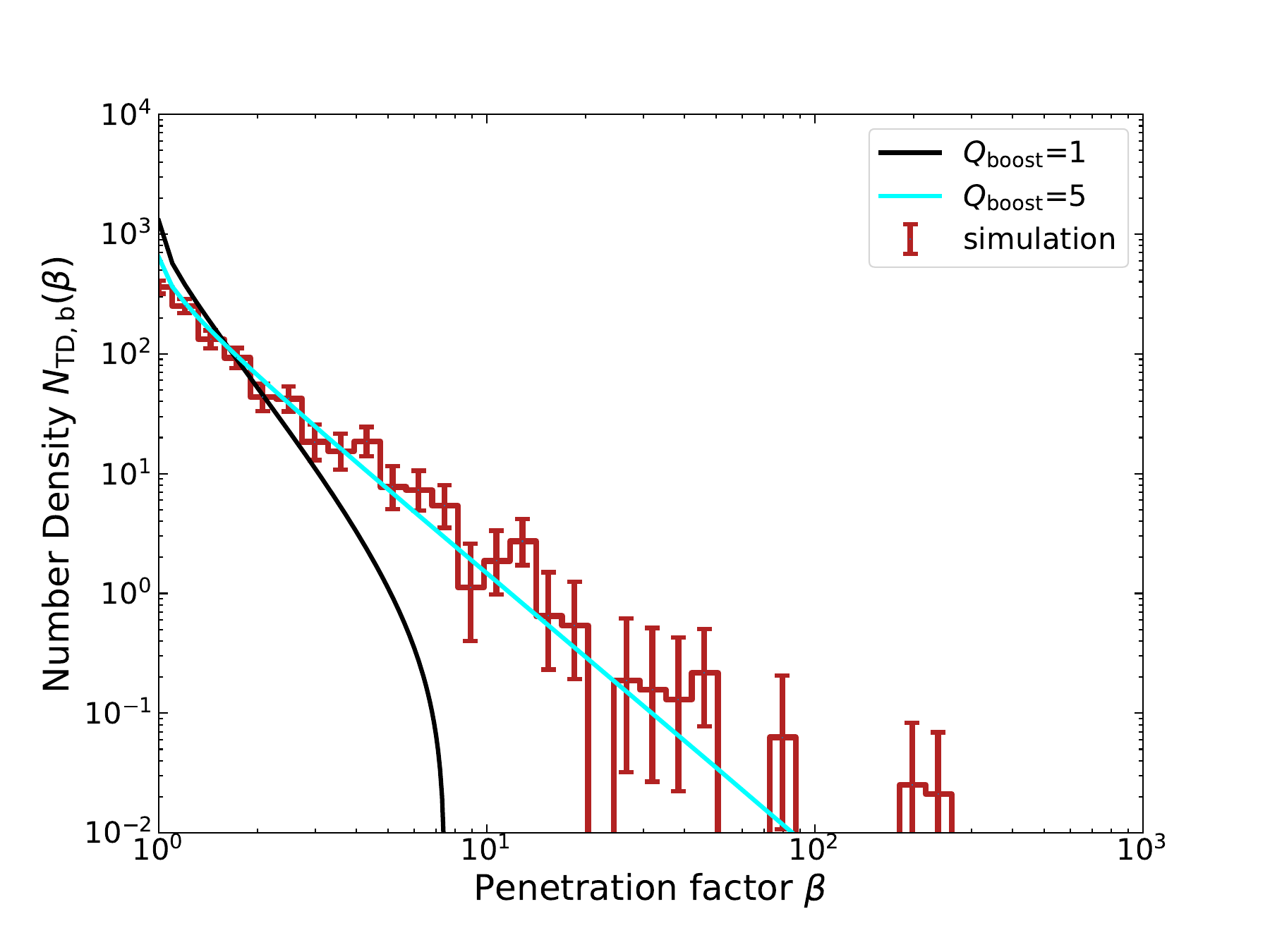}
  \includegraphics[width=0.47\columnwidth]{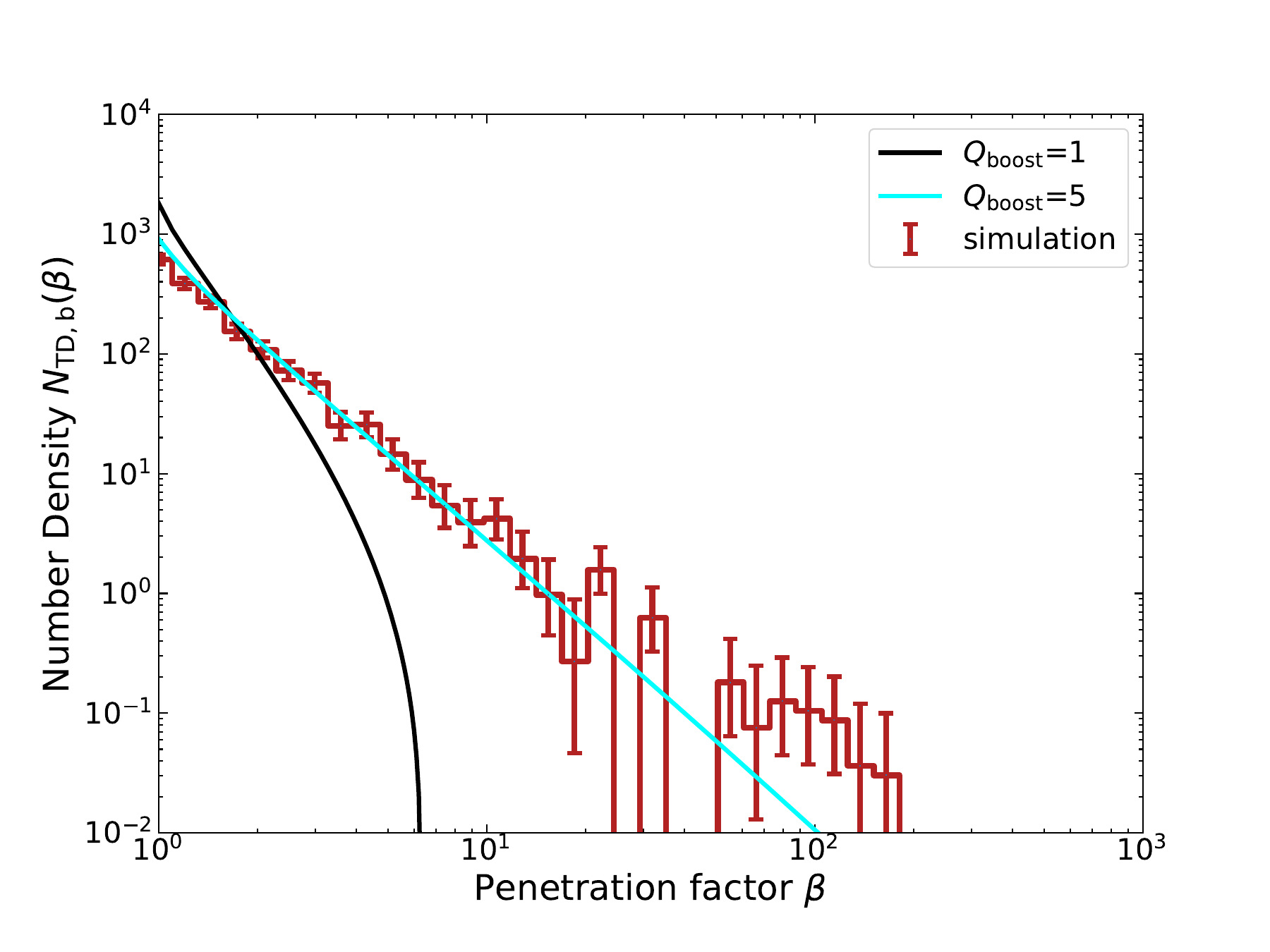}
  \end{center}
\caption{
Dependence of the number density of the bound stars on the penetration factor ($\beta$). The figure formats are the same as Figure~\ref{fig_n_e_b} but for $\beta$. 
}
\label{fig_n_b_b}
\end{figure}

In the previous subsection, we have derived theoretical, analytical expressions for the distribution of tidally accreted stars (bound and unbound ones) in terms of eccentricity $e$ and penetration factor $\beta$; in order to achieve that, we have used double power law functions for the stellar density, the Jeans equation for the stellar velocity dispersion, and double-power-law function for the energy distribution of tidally disrupted stars.
Now we will check the final results of the previous subsection for
$N_{\rm TD,b,u}(e)$ and $N_{\rm TD,b,u}(\beta) $ (see equation~\ref{Eq-n(beta)-bound}) directly against the $N$-body data of particles arriving at the tidal radius.
Figure~\ref{fig_n_e_b} shows the dependence of the number density of bound stars on the orbital eccentricity.
The red histogram represents the simulated number densities. The uncertainties of the measurements are computed based on the Poisson error, the $1\sigma$ confidence level single-sided upper and lower limits are computed with equations (9) and (12) in \citet{Gehrels1986}. The black and cyan curves represent the theoretical number densities obtained with different $Q_{\rm boost}$ values (note this specific color setting for $Q_{\rm boost}$ is used in Figures~\ref{fig_Q-r_a},~\ref{fig_n_e_b},~\ref{fig_n_b_b} and \ref{fig_boundary}). We find that $Q_{\rm boost}$ can mildly affect the theoretical $N_{\rm TD,b}(e)$. In both panels, the distributions are quite narrow near the parabolic case ($e=1$). The simulated number densities are also in good agreement with the theoretical ones, except for some stronger fluctuations around $e \sim 0.996$ and $ e \sim 0.998$. This is because the particle resolution of our $N$-body simulations is not sufficient there.
The number density of the $M_{\rm BH}=0.01$ case is wider for the orbital eccentricity than that of the $M_{\rm BH}=0.05$ case. This trend can be interpreted as follows: from equation (\ref{eq:ell}), the lowest eccentricity
of the bound stars can be estimated to be $e_{\rm ll}(M_{\rm BH},E_{\rm min})
= \sqrt{1-4r_{\rm t}|E_{\rm min}|/GM_{\rm BH}}$. We find that $E_{\rm min}=-2$
in the $M_{\rm BH}=0.01$ case, whereas $E_{\rm min}=-4$ (ignoring the isolated bins)
in the $M_{\rm BH}=0.05$ case. Substituting each quantity into the above equation,
we find $e_{\rm ll}(0.01,-2)=0.9960$ and $e_{\rm ll}(0.05,-4)=0.9984$.
These evaluations are consistent with the number density distributions shown
in Figure~\ref{fig_n_e_b}.
%

Figure~\ref{fig_n_b_b} shows the dependence of the number density of bound stars on
the penetration factor $\beta$. The figure format of the two panels is the same as for Figure~\ref{fig_n_e_b}. 
We find $Q_{\rm boost}$ has strong effect on the theoretical $N_{\rm TD,b}(\beta)$. For bound stars, the maximum value of $Q$ is achieved at $r_a = r_{\rm h}$. Figure~\ref{fig_Q-r_a} shows that in the  $M_{\rm BH}=0.01$ case, 
$Q(r_{\rm h})=(1.89,~9.45)$ for $Q_{\rm boost}=(1,~5)$, respectively.
Since $N_{\rm TD,b}(\beta,E)=0$ when $\beta>1/g(Q)$, the integrated $N_{\rm TD,b}(\beta)$ shall vanish beyond $\beta=7.4$ in the $Q_{\rm boost}=1$ case, while for the other $Q_{\rm boost}$ case, the vanishing point extends to much higher $\beta$, as shown in the left panel of Figure~\ref{fig_n_b_b} [the vanishing behavior of the theoretical models in the right panel ($M_{\rm BH}=0.05$) can be understood in the same way].
We find that the analytical number densities obtained with $Q_{\rm boost}=5$ are in good agreement with the simulated ones within the range of $\beta\lesssim10$. For $\beta\gtrsim10$, the deviation between the analytical and simulated number densities gets larger, because of the poor numerical resolution of the N-body models.
%
%
\begin{figure}[htbp]
  \begin{center}
  \includegraphics[width=0.47\columnwidth]{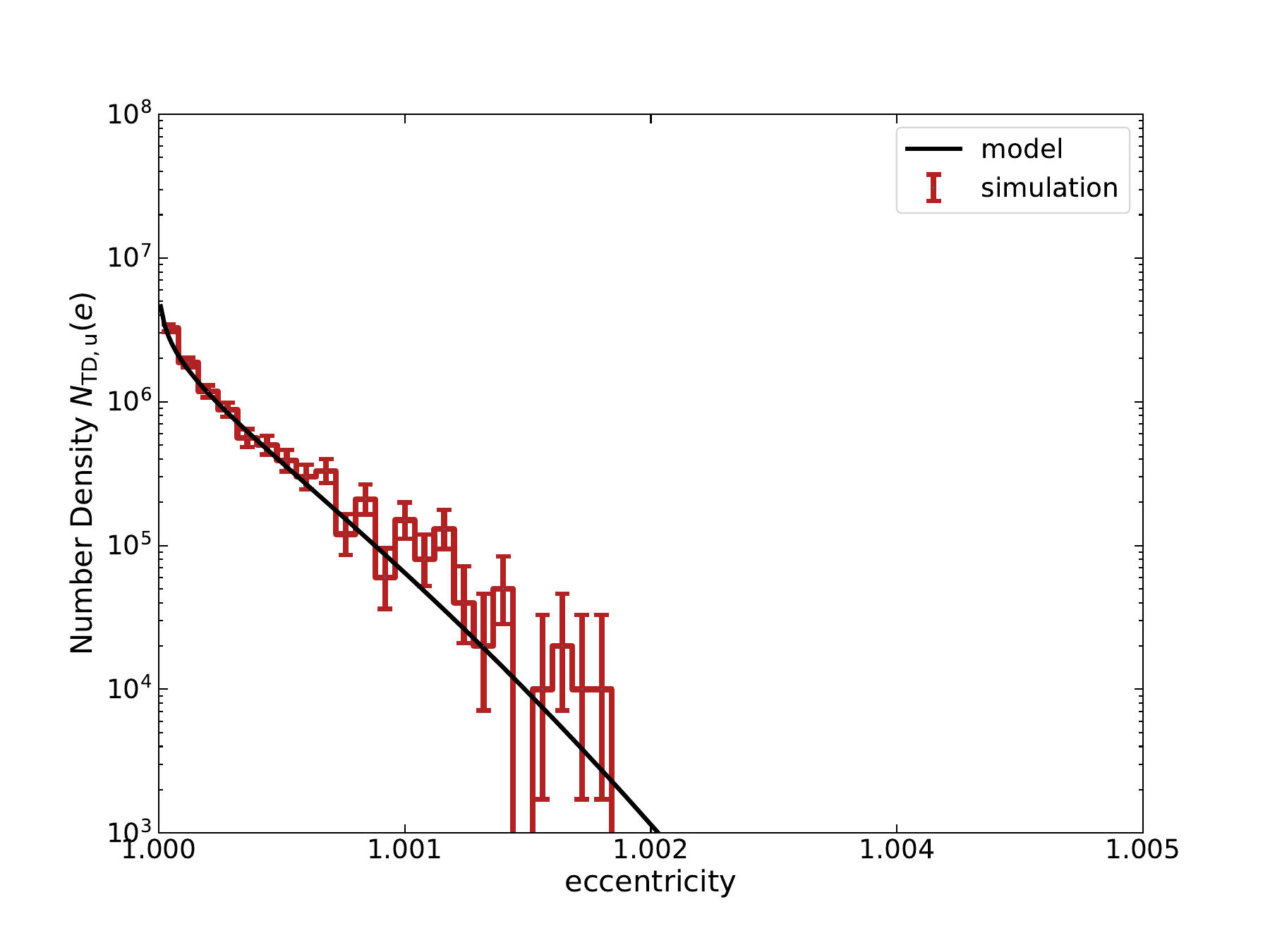}
  \includegraphics[width=0.47\columnwidth]{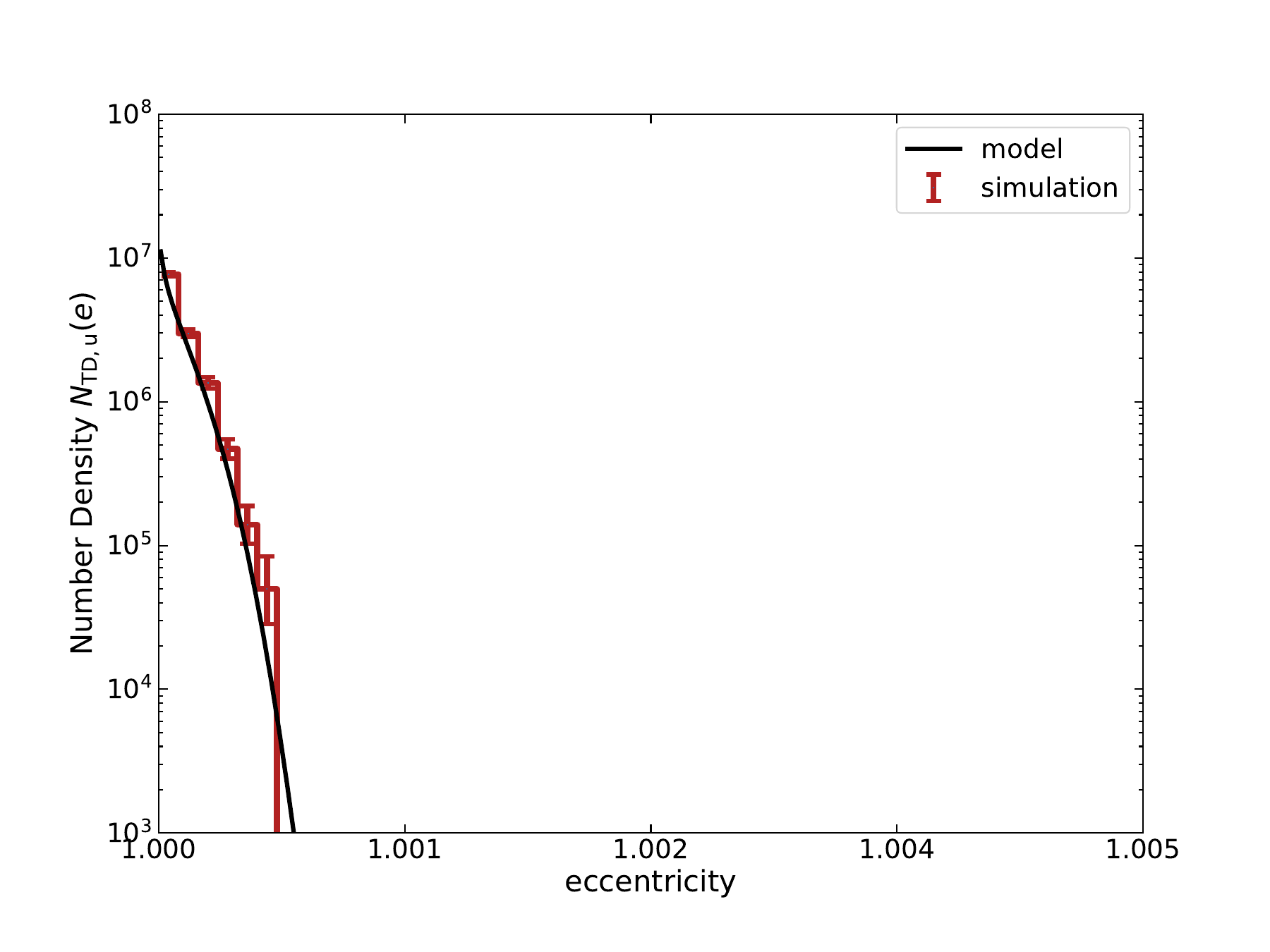}
  \end{center}
  \caption{Same format as Figure~\ref{fig_n_e_b} but for the unbound stars. Equation~\ref{Eq-n(e,E)-unbound} is used to evaluate the theoretical number density quantitatively.
  }
  \label{fig_n_e_u}
\end{figure}

%
%
\begin{figure}[htbp]
  \begin{center}
  \includegraphics[width=0.47\columnwidth]{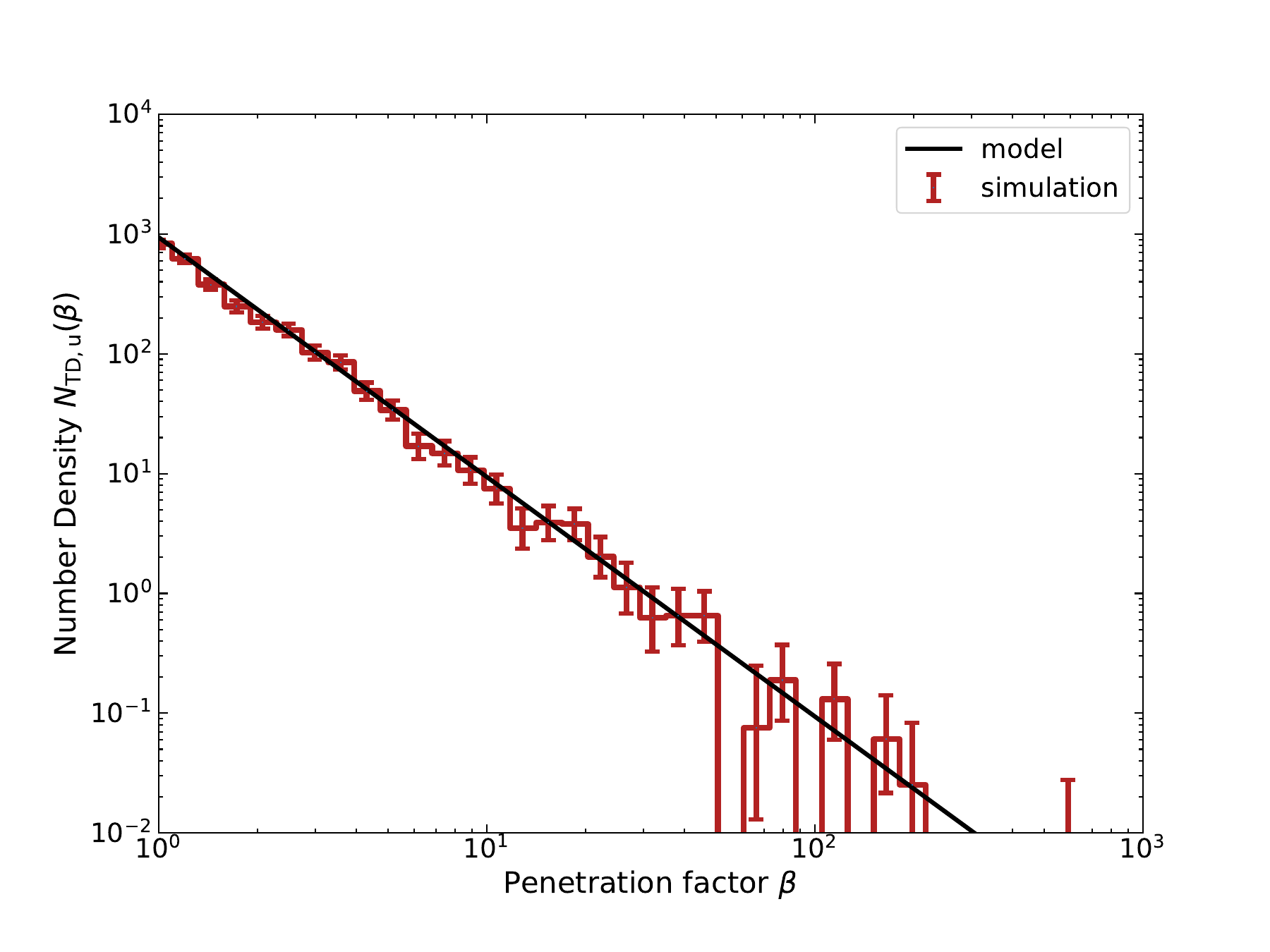}
  \includegraphics[width=0.47\columnwidth]{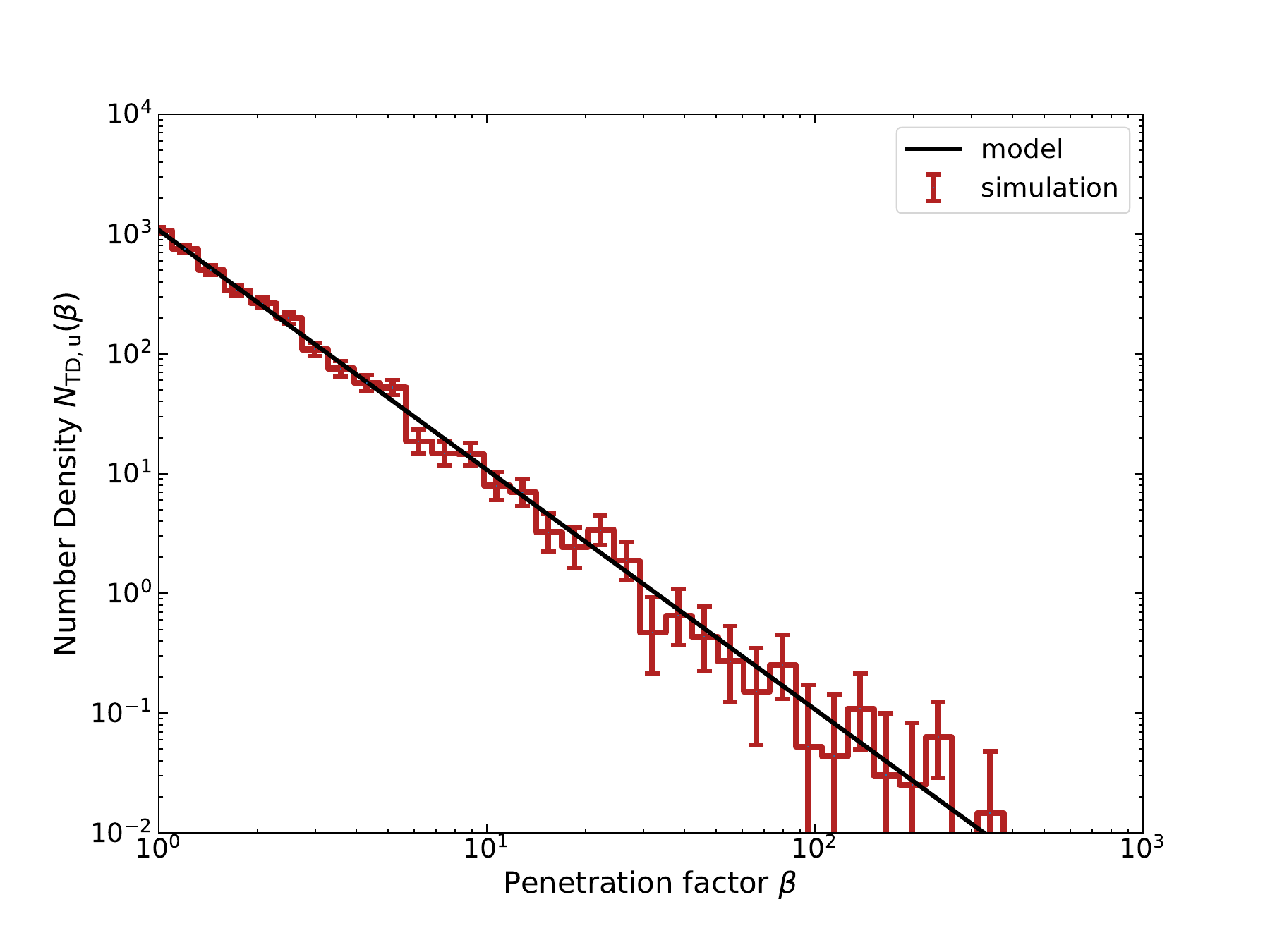}
  \end{center}
  \caption{Same format as Figure~\ref{fig_n_b_b} but for the unbound stars. Equation~\ref{Eq-n(beta,E)-unbound} is used to evaluate the theoretical number density quantitatively.
  }
  \label{fig_n_b_u}
\end{figure}

Figures~\ref{fig_n_e_u} and \ref{fig_n_b_u} compare the analytical number densities of
$N_{\rm TD,u}(e)$ and $N_{\rm TD,u}(\beta)$ with those of the simulated number densities.
Our theoretical predictions match well with the simulated number densities in both figures.
As in the bound star case, the number densities of $M_{\rm BH}=0.01$ case are more widely
distributed over the eccentricity than the $M_{\rm BH}=0.05$ case.
Substituting $E_{\rm max}$ and $\vert E_{\rm t}\vert = GM_{\rm bh}/r_{\rm t}$ into equation~(\ref{Eq-beta-e-unbound})
with $\beta = 1$, we obtain $e_{\rm max}(M_{\rm BH},E_{\rm max}) = 1 + 2r_{\rm t}E_{\rm max}/(GM_{\rm BH})$.
We find from the simulated value of $E_{\rm max}$ that $e_{\rm max}(0.01,1.259)=1.0025$ and
$e_{\rm max}(0.05,1.585)=1.0006$. These suggest that the number density is more widely
distributed over the orbital eccentricity in the star cluster with the less massive black hole.
%

%
\subsection{Distribution of stars on the eccentricity-penetration factor plane}
\label{sub:dis}
%
At the time of tidal disruption, it is the eccentricity $e$ and the penetration factor $\beta$ that can be related to the observational characteristics (e.g., light curve). Also in our $N$-body simulations, we have a direct handle to determine these two quantities for any tidal disruption locally, without knowing anything about the large-scale distribution of stars and the gravitational potential. Therefore, we check here what we can deduce from our previous analytical results for the distribution of tidally disrupted stars on the $e$-$\beta$ plane and compare again the expectations with the simulation data.
For a given energy $E=-GM_{\rm BH}/2a$ at the $\mathcal{R}=\mathcal{R}_0$ limit, we can define the minimum orbital eccentricity of a bound star by using $r_{\rm t}/\beta=a(1-e)$ as
\begin{equation}
e_{\rm min} = 1-\frac{2}{\beta}\frac{r_{\rm t}|E|}{GM_{\rm BH}},
\label{Eq-b-boundary}
\end{equation}
where $\beta=1/g(Q)$ is obtained through equations~(\ref{Eq-R0}) and (\ref{Eq-beta}).
Adopting $\beta=1$, we find equation~(\ref{Eq-b-boundary}) corresponds to equation~(\ref{eq:ell})
at $r_{\rm t}/a\ll1$: $e_{\rm ll}=\sqrt{1-4r_{\rm t}|E|/(GM_{\rm BH}})\approx1-2r_{\rm t}|E|/(GM_{\rm BH})$.
The unbound stars all have $E \leq E_{\rm max}$.
We use again  $\vert E_{\rm t}\vert = GM_{\rm bh}/r_{\rm t}$ and equation~(\ref{Eq-beta-e-unbound}), to obtain
\begin{equation}
e_{\rm max} = 1 + \frac{2}{\beta}\frac{r_{\rm t}E_{\rm max}}{GM_{\rm BH}}.
\label{Eq-u-boundary}
\end{equation}
For $Q=1$, we obtain the orbital eccentricity on the boundary between the empty loss-cone and the full loss-cone regimes:
\begin{equation}
e_{\rm lcb} = 1 - \frac{r_{\rm t}}{a_{\rm lcb} \beta},
\label{Eq-Q=1}
\end{equation}
where the corresponding semimajor axis, $a_{\rm lcb}$, can be obtained from Eq.~\ref{eq:appq} as
\begin{equation}
a_{\rm lcb}=Q_{\rm boost}\left(\frac{t_{\rm relax}}{t_{\rm dyn}}\right)r_{\rm t} \ ,
\end{equation}
where we have adopted $r_{\rm a}=2a$, and $r_{\rm a}$ is computed from the combined gravitational potential of stars and SMBH, see right panel of Figure~\ref{fig_sig-rho} for the result.
In this case, we find $e_{\rm lcb}\approx1$ for $\beta\ge1$.

\begin{figure}[htbp]
  \begin{center}
  \includegraphics[width=0.47\columnwidth]{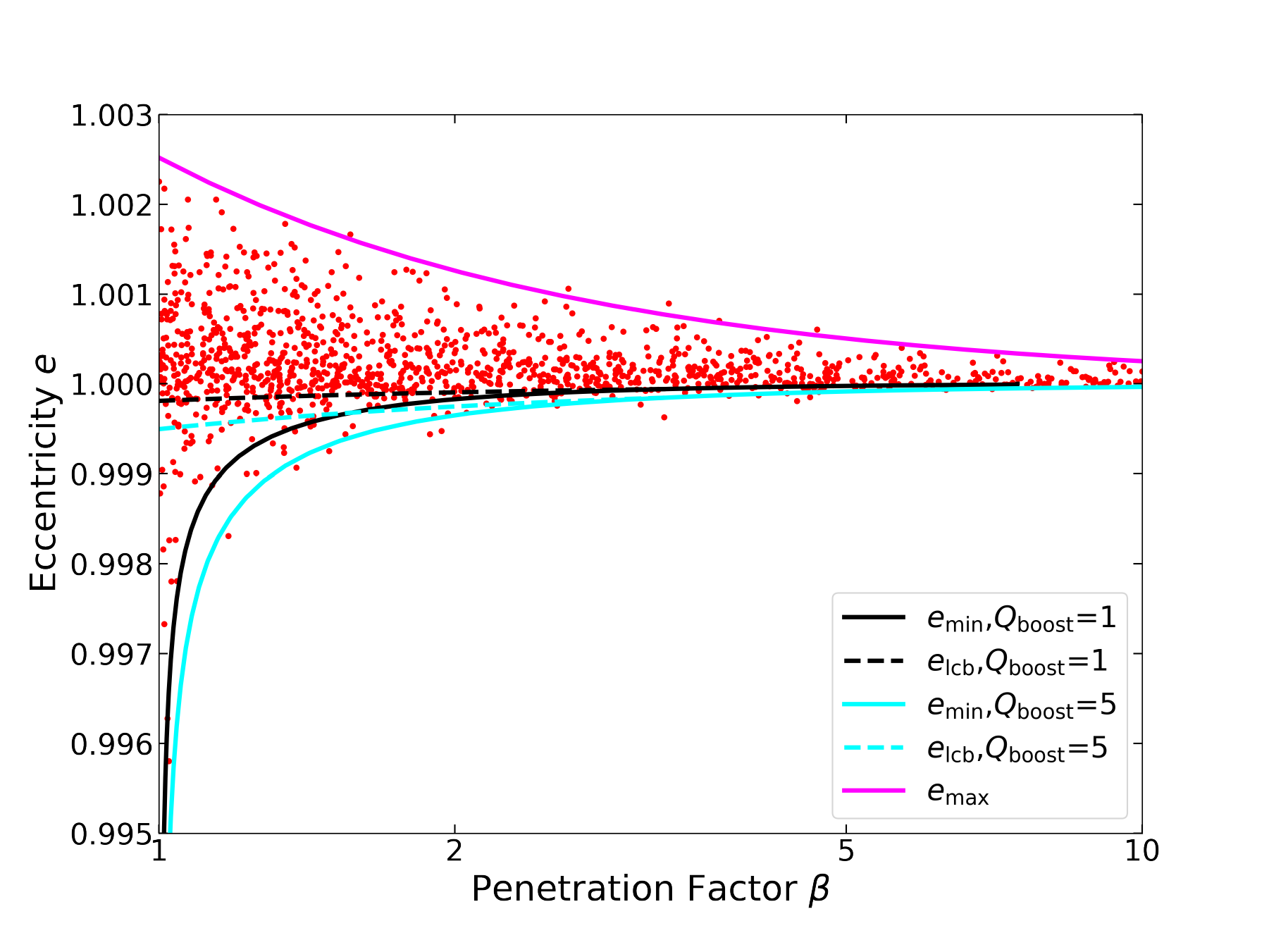}
  \includegraphics[width=0.47\columnwidth]{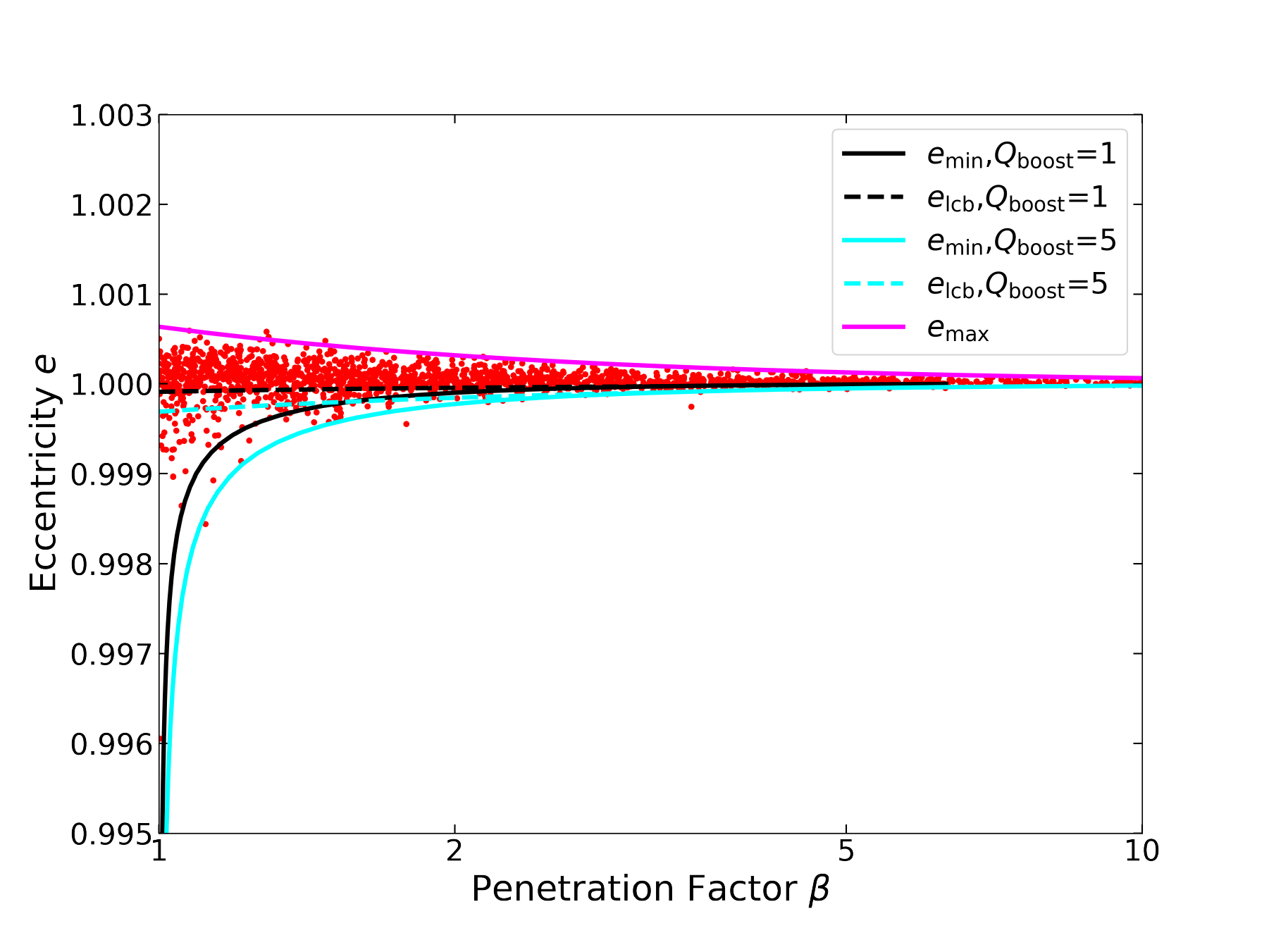}
  \end{center}
\caption{
Distribution of the stars, which can cause TDEs, on the $e$-$\beta$ plane
(left: $M_{\rm BH}=0.01$; right: $M_{\rm BH}=0.05)$.
The solid magenta curve denote $e_{\rm max}$. The $e_{\rm min}$ (solid) and $e_{\rm lcb}$ (dashed) are plotted with two different $Q_{\rm boost}$ values that are indicated by the colors, respectively (see also equations~\ref{Eq-b-boundary}-\ref{Eq-Q=1}).
}
\label{fig_boundary}
\end{figure}

Figure~\ref{fig_boundary} shows the distribution of bound and unbound stars that undergo TDEs on the $e$--$\beta$ plane. 
The solid magenta curve denote $e_{\rm max}$. The $e_{\rm min}$ (solid) and $e_{\rm lcb}$ (dashed) lines are plotted with two different $Q_{\rm boost}$ values that are indicated by the colors, respectively. The value of $E_{\rm max}$ is obtained from our $N$-body simulations and $a_{\rm lcb}\approx r_{\rm crit}/2$ is read out from Figure~\ref{fig_Q-r_a}. We notice that $Q_{\rm boost}=1$ gives poor estimate of $e_{\rm min}$: lots of data points are lying outside the boundary. On the other hand, the $Q_{\rm boost}=5$ models show better agreement with the data. While the space between $e_{\rm lcb}$ and $e_{\rm min}$ corresponds to the empty loss-cone regime, the space between $e_{\rm lcb}$ and $e=1$ corresponds to the full loss-cone regime. The stars located above the $e=1$ axis are supplied to the black hole from the Maxwellian distribution regime. We find that only a small fraction of the stars originate from the empty loss-cone regime and the corresponding values of $\beta$ are distributed around unity, whereas most of the stars are originally supplied from the region between $e_{\rm lcb}$ and $e_{\rm max}$ over the much wider range of $\beta$. These results are consistent with loss-cone theory: due to the diffusion nature in the empty loss-cone regime, the loss-cone flux is much smaller than the full loss-cone regime, and the stars inside the empty loss-cone cannot penetrate the tidal radius too much.

We also find that there are some stars outside $e_{\rm min}$ or $e_{\rm max}$. These outliers seem to violate the loss-cone theory. There are two possible reasons for the outliers behaving unexpectedly: first, the more energetic close two-body encounter
occurs in the $N$-body simulations, leading to an enhancement of the angular momentum exchange, so that the stars have $\mathcal{R} < \mathcal{R}_0$ \citep{LT1980}. Second,
the quantity $Q$ depends on the density and the velocity dispersion of the star cluster, which can fluctuate with radius and evolve with time. The number of outliers is very small, so $e_{\rm min}$ and $e_{\rm max}$ are generally useful as limits.

Note that there are two critical eccentricities connected to a TDE, $e_{\rm crit,1}$ and $e_{\rm crit,2}$, which divide eccentricity space (and TDEs) into five regimes: 1 - eccentric (1: $0\!<e\!<e_{\rm crit,1}$); 2 - marginally
eccentric ($e_{\rm crit,1}\!\le\!e\!<\!1$); 3 - purely parabolic ($e\!=\!1$); 4 - marginally hyperbolic
($1\!<\!e\!<\!e_{\rm crit,2}$); and 5 -- hyperbolic ($e\!\ge\!e_{\rm crit,2}$), respectively. Assuming some model of the internal structure of the disrupted star, \cite{HZL2018} and \cite{2020ApJ...900....3P} find that in regime 1 the stellar debris fully accreted, in regime 5 all the debris escapes, while in the regimes 3, 4, and 5 we have the partial accretion and escape of the debris. Regime 4 refers to the standard case of \cite{R1988} with half of the debris escaping.

The critical eccentricities are \citep{2020ApJ...900....3P}:
\begin{equation}
e_{\rm crit,1}=1-2(M_{\rm bh}/m_*)^{-1/3}\beta^{k-1} \ \ ; \ \ 
e_{\rm crit,2}=1+2(M_{\rm bh}/m_*)^{-1/3}\beta^{k-1} \ \ ,
\end{equation}
where $k=0$ is the original case discussed in \cite{HZL2018}.
They found by $N$-body experiments that stars on marginally eccentric and marginally hyperbolic orbits are the main source of TDEs in a spherical nuclear star cluster. 

In our case, we have $m_\star = 1/N$; for the simulations used in Figure~\ref{fig_boundary} we find, for example,
$e_{\rm crit,1}\approx0.885$ and $e_{\rm crit,2}\approx1.12$ for
$M_{\rm bh}=0.01$, $\beta=1$, and $k=0$ case and $e_{\rm crit,1}\approx0.933$
and $e_{\rm crit,2}\approx1.067$ for $M_{\rm bh}=0.05$, $\beta=1$, and $k=0$ case. 
In any case, the maximum and minimum eccentricities predicted and found in the simulation, as shown in the figure, are much smaller than $e_{\rm crit,2}$ and much larger than $e_{\rm crit,1}$, i.e. very close to the parabolic case. This means, in the terminology of \cite{HZL2018} that all our TDEs are either marginally eccentric or marginally hyperbolic. This is not surprising, because our analysis is based on the same $N$-body data; however, we have derived in this subsection much narrower limits for $e_{\rm min}$ and $e_{\rm max}$, which are based on the diffusion theory going back to CK78.

%
\section{Discussion}
\label{SECT-DISCUSSION}
%
We have derived a semi-analytical model for the number distribution of eccentricity $e$ and penetration factor $\beta$ of tidally disrupted stars in a spherical nuclear star cluster around an SMBH. It has been compared to the results of our previously published \citep{HZL2018} direct $N$-body simulation with good agreement. To get our model, 
we use double-power-law functions to fit the stellar density of the $N$-body data and compute the velocity dispersion profile via the Jeans equation, as well as use double-power-law function to model the energy distribution of the bound and unbound tidally disrupted stars.
Our method is based on the classical results of loss-cone diffusion for bound stars, using the Fokker-Planck equation (CK78). For unbound stars, we use a simple approximation based on the assumed Maxwellian character of the stellar distribution function.
For an improved treatment of unbound stars, we need to consider the effects of the local self-gravity of stars and other external factors as the galactic potential. 
Our model is useful for discussing the scaling behavior of TDE statistics; current $N$-body simulations are still far away from realistic particle numbers and sizes of tidal radii \citep{HZL2018}. Nevertheless, simulation data like the ones presented here have been used to extrapolate from our unphysically small particle numbers and large tidal disruption radii to real galactic nuclei - typical results for the TDE rate in $N$-body simulation models range around or little above $10^{-6}$ per year per galaxy \citep{ZBS2014,Panamarev2019,LZB2023}, which is the lower bound of \cite{SM2016}; the latter paper gives a range of up to $10^{-4}$ per year per galaxy, in accord with recent work of \cite{Bortolas2023}. When comparing such rates, one should bear in mind that the goal of our paper is not to give any accurate predictions of observed TDE rates. Cited papers include stars from a much wider range of origin, out into the bulge; our work has not yet properly accounted for a realistic mass spectrum and tidal disruption properties depending on stellar type and parameters (but this work is in progress). Another parameter not yet carefully checked in our models is the effect of the black hole mass (relative to the cluster mass and relative to the stellar particle mass). 

\begin{figure}[!htbp]
\centering
\includegraphics[width=0.9\columnwidth]{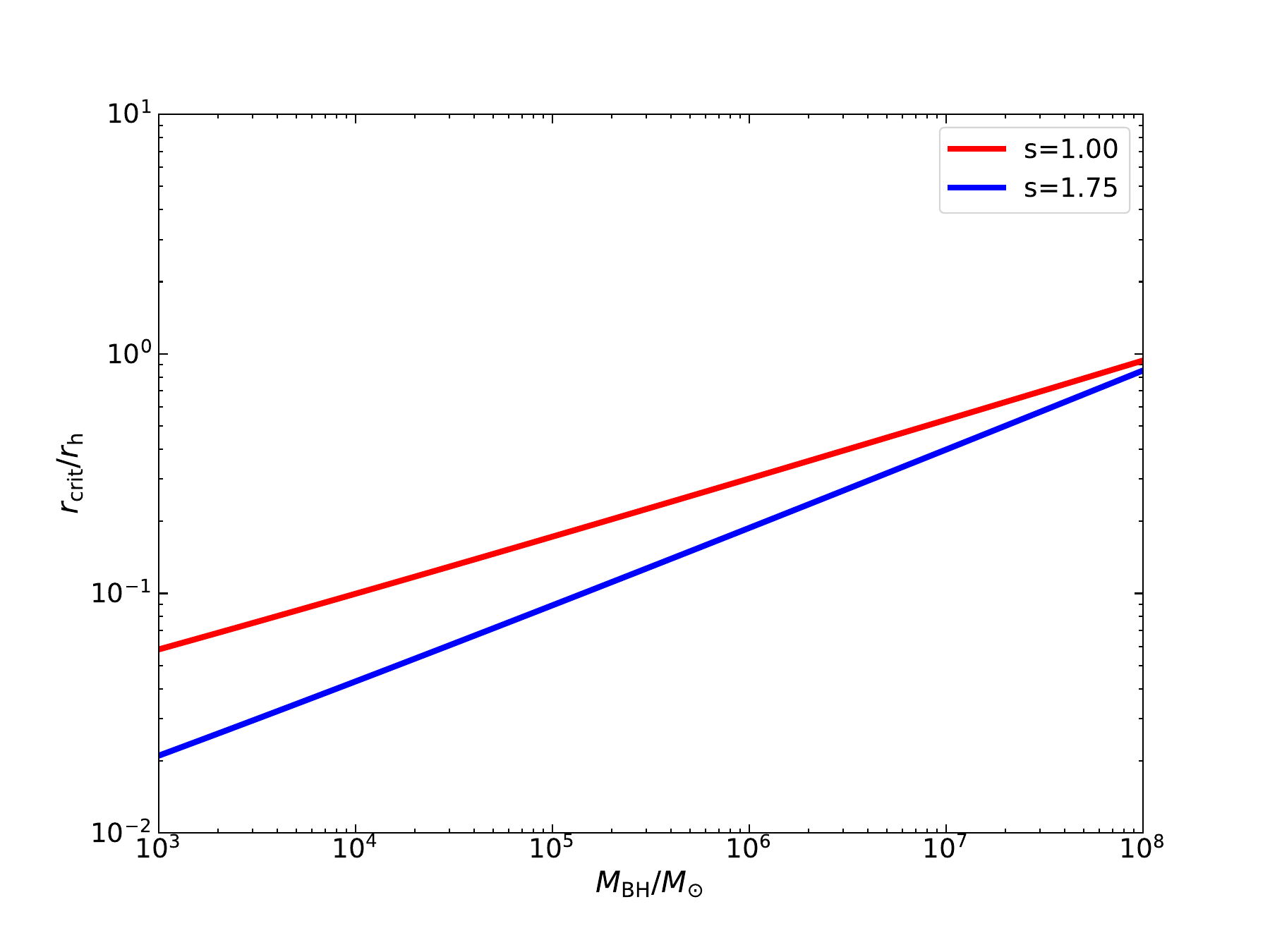}
\caption{
Black hole mass dependence of $r_{\rm crit}/r_{\rm h}$.
The two slopes for the density profiles: $s=1.75$
(Bahcall--Wolf cusp) and $s=1$ ($N$-body simulations)
are adopted.
}
  \label{fig_rcrit-MBH}
\end{figure}

Let us first qualitatively discuss how the number densities derived in our work would vary with the black hole mass relative to the star cluster. We assume a power-law density profile
in the cusp $\rho(r) = \rho_h(r/r_{\rm h})^{-s}$, where $\rho_{\rm h}=(3-s)M_{\rm BH}/(4\pi r_{\rm h}^3)$ is the density at the influence radius according to the definition of $r_{\rm h}$. Note that the total stellar mass inside the influence radius is equal to the black hole mass. The velocity dispersion inside the influence radius follows $\sigma^2(r) \approx G M_{\rm BH}/r$. A critical radius, where the star consumption is balanced with its replenishment by two-body relaxation, is defined by the conditions $\theta_{\rm lc} = \theta_{\rm D}$ (in the notation of \cite{FR1976} and \cite{2004MNRAS.352..655A}) or $Q=1$ (in our notation following CK78). Following \cite{FR1976} and \cite{BBK2011} we get
\begin{equation}
r_{\rm crit} = \Bigl [ \frac{2 M^2_{\rm BH}r_{\rm t}N}{2.94 Q_{\rm boost} \ln(0.11N) M_{\rm c} \rho_{\rm h} r^s_{\rm h}} \Bigr ]^{\frac{1}{4-s}}\propto M_{\rm BH}^{\frac{17-3s}{6(4-s)}}
\label{Eq-rcrit}
\end{equation}
As we have shown in the previous section, a $Q_{\rm boost}$ factor is important for matching the theoretical model with the $N$-body results, so we also add the $Q_{\rm boost}$ factor to equation~\ref{Eq-rcrit} and in the following part we adopt $Q_{\rm boost}=5$. Note that the left term yields for the traditional value $s=7/4$ the result $r_{\rm crit}\propto (r_{\rm t} M_{\rm BH})^{(4/9)}$ consistent with \cite{BME2004a}. The right-hand side delivers a different scaling, because we have used additionally results of the scaling procedure described in \citet{ZBS2014}: $r_{\rm t} \propto M^{1/3}_{\rm BH}$, $N \propto M_{\rm BH}$,
$M_{c} \propto M_{\rm BH}$, $r_{\rm h} \propto M^{1/2}_{\rm BH}$,
and $\rho_{\rm h} \propto M_{\rm BH}/r^3_{\rm h} \propto M^{-1/2}_{\rm BH}$, and we simply adopt $1/2$ for the scaling of $r_{\rm h}$ instead of $0.54$, which is obtained from the $M_{\rm BH}-\sigma$ relation, $\log(M_{\rm BH}/M_{\odot})=8.18+4.32\log[\sigma/(200 \rm{km~s^{-1}})]$ \citep{SG2011}. We get for the ratio of critical to influence radius using our scaling
\begin{equation}
\frac{r_{\rm crit}}{r_{\rm h}} = \left[\frac{2}{2.94}
\frac{4\pi}{(3-s)}\frac{1}{Q_{\rm boost}}\frac{M_{\rm BH}}{M_{\rm c}}\frac{N}{\ln(0.11N)}
\frac{r_{\rm t}}{r_{\rm h}}
\right]^{\frac{1}{4-s}}
\label{Eq-rcrit_over_rh}
\end{equation}
is proportional to $M_{\rm BH}^{5/[6(4-s)]}$ (ignoring the slowly varying logarithmic term).
Therefore, for $s<4$, the ratio of critical to influence radius increases with black hole mass, as was already observed for the standard case by \cite{FR1976}. Figure~\ref{fig_rcrit-MBH} depicts $r_{\rm crit}/r_{\rm h}$ as a function of $M_{\rm BH}$ (see also equation~\ref{Eq-rcrit_over_rh}) in the range of $10^3\,M_\odot\le{M_{\rm BH}}\le10^8\,M_\odot$, where we adopt $s=1.75$ for the Bahcall--Wolf cusp \citep{BW1976} and $s=1$ for the cusp obtained from the $N$-body simulations.

\begin{figure}[htbp]
\begin{center}
\includegraphics[width=0.47\columnwidth]{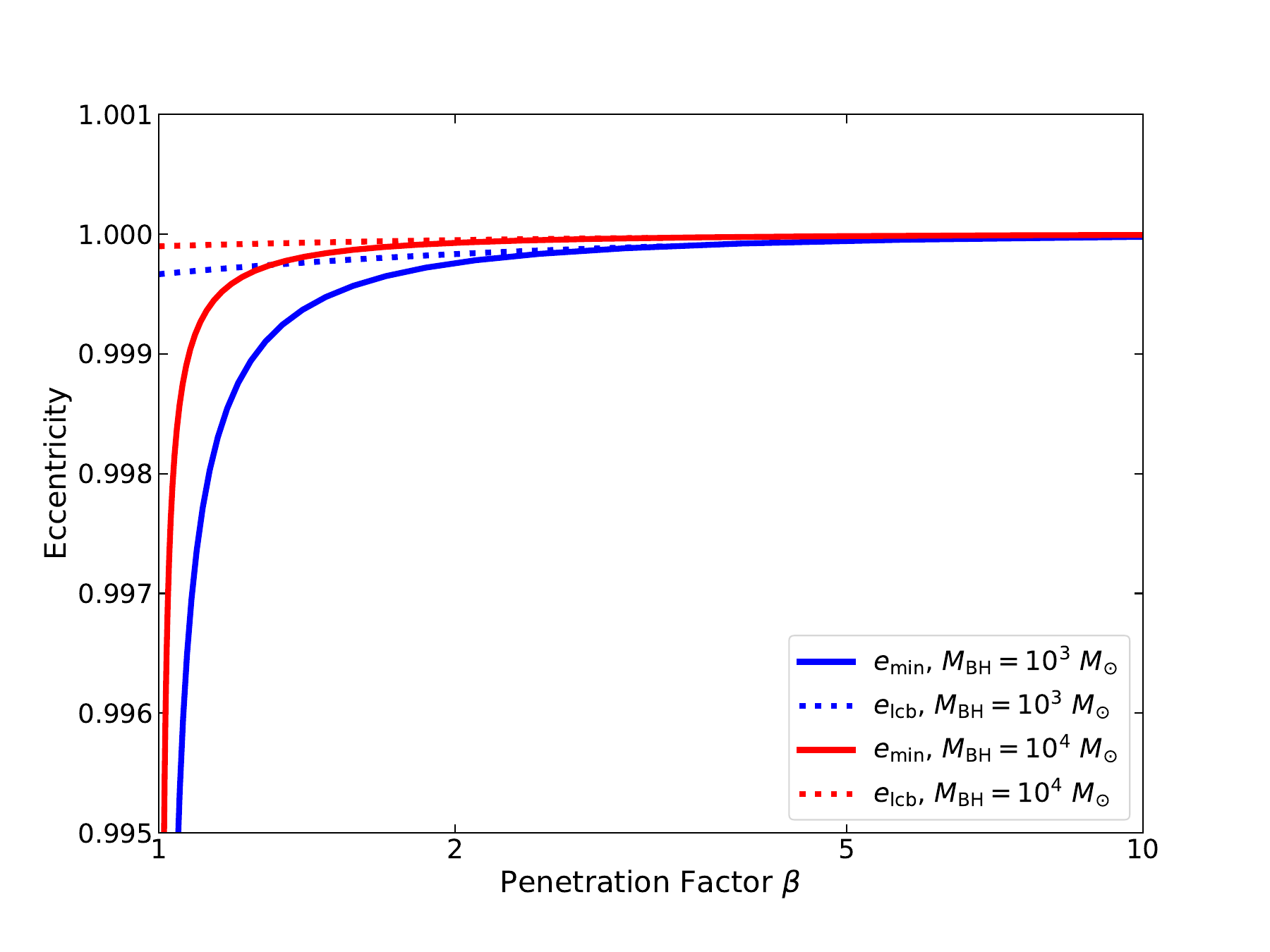}
\includegraphics[width=0.47\columnwidth]{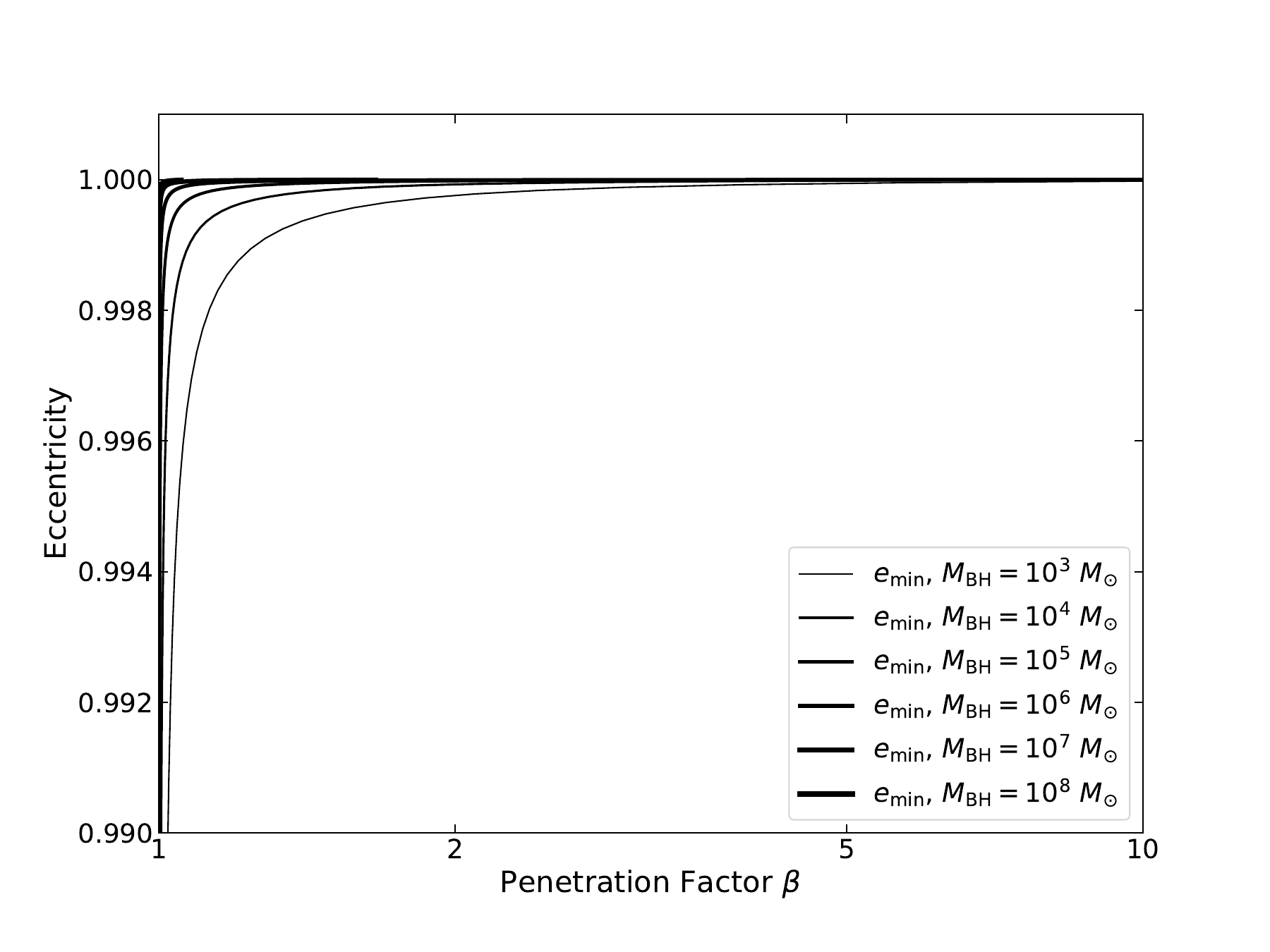}
\end{center}
\caption{
Black hole mass dependence of two characteristic eccentricities: $e_{\rm min}$ and $e_{\rm lcb}$
in the $e$-$\beta$ plane. We adopt the Bahcall--Wolf density cusp and $Q_{\rm boost}=5$ for all the models.
The left panel shows the $\beta$-dependence of $e_{\rm min}$ and $e_{\rm lcb}$ for the
$M_{\rm BH}=10^3 M_{\odot}$ and $10^4 M_{\odot}$ cases. The dashed lines denote the eccentricity,
$e_{\rm lcb}$, between the empty and full loss-cone regimes (see equation~\ref{Eq-Q=1}), whereas the
solid lines denote the eccentricity, $e_{\rm min}$, of most tightly bound stars (see equation~\ref{Eq-emin}).
The different color shows the different black hole mass. In the right panel, all the lines represent $e_{\rm min}$.
We adopt different line styles for the different black hole masses for $10^3\,M_\odot\le{M}_{\rm BH}\le10^8{M}_{\odot}$.
}
\label{fig_beta-limit-MBH}
\end{figure}

%
Assuming that the semimajor axis $a_{\rm min}$ corresponding to $E_{\rm min}$
is given by a fixed fraction of $r_{\rm crit}$ as $a_{\rm min}=fr_{\rm crit}$,
where $f$ is a parameter determined by $N$-body simulations, we have
$E_{\rm min} \propto -M_{\rm BH}^{\frac{7-3s}{6(4-s)}}$.
When $s < 7/3$, $E_{\rm min}$ decreases as $M_{\rm BH}$ decreases.
Substituting $E_{\rm min} = -GM_{\rm BH}/(2f r_{\rm crit})$ into equation~(\ref{Eq-b-boundary}), we obtain
\begin{equation}
e_{\rm min}=1-\frac{1}{\beta}\frac{1}{f}\frac{r_{\rm t}}{r_{\rm crit}}.
\label{Eq-emin}
\end{equation}
Since the ratio of $r_{\rm t}$ to $r_{\rm crit}$ is less than $3\times10^{-4}$
over the whole range of the black hole mass, $e_{\rm min}$ is always larger: $0.994$ for $f=0.05$ and $\beta=1$. Because of $1-e_{\rm min}\propto{r_{\rm t}/r_{\rm crit}}\propto{M_{\rm BH}^{(s-9)/[6(4-s)]}}$, $e_{\rm min}$ is closer to 1 for $s<4$ as the black hole mass is larger.
We also confirm that the black hole mass dependence of the pericenter radius
is the same as that of the tidal disruption radius, i.e.,
$r_{p,\rm {min}}=a_{\rm min}(1-e_{\rm min})\propto{M_{\rm BH}^{1/3}}\propto{r_{\rm t}}$.
Next, let us see how $e_{\rm min}$ and $e_{\rm lcb}$ (see equations \ref{Eq-Q=1}
and \ref{Eq-emin}) depend on the black hole mass on the $e$-$\beta$ plane. The left
panel of Figure~\ref{fig_beta-limit-MBH} shows it for the $M_{\rm BH} = 10^3$ and
$10^4 M_{\odot}$ cases. The $e_{\rm lcb}$ curve of $M_{\rm BH} = 10^3 M_{\odot}$
gets larger than $M_{\rm BH} = 10^4 M_{\odot}$ case (thick curve). In addition, as
mentioned in Section~\ref{sub:dis} (see also Figure~\ref{fig_boundary}), most of the
bound stars are distributed around the $e_{\rm lcb}$ curve on the plane. These suggest
that the star is distributed closer to $e=1$ over the whole range of $\beta$ as the black
hole mass increases. The right panel of Fig.~\ref{fig_beta-limit-MBH} depicts how the
$e_{\rm min}$ curve depends on the black hole mass on the $e$-$\beta$ plane. It is clear
from the panel that $e_{\rm min}$ gets closer to 1 as the black hole mass is larger.
This is consistent with $1-e_{\rm min}\propto M_{\rm BH}^{(s-9)/[6(4-s)]}\,(s<4)$
as we estimated in the previous paragraph. In summary, Figure~\ref{fig_beta-limit-MBH}
suggests that the stars are supplied into the black hole on extremely marginally eccentric
to parabolic orbits for a spherical cluster with $M_{\rm BH} > 10^7 M_{\odot}$ black hole.

Our analysis of the $e$-$\beta$ distributions could provide a good tool to probe the
dynamical status of the stars causing TDEs in a star cluster. \citet{HSL2013, 2016MNRAS.461.3760H}
and \citet{BRL2016} studied the accretion disk formation by performing hydrodynamic
simulations, where the authors have adopted $(\beta,e)=(5,0.8)$ as initial values, although
they also used other combinations of $(e,\beta)$.
This parameter set is clearly ruled out in our model. However, this does not mean
that such a very tightly bound TDE cannot occur. The tightly bound stars are likely to
supply to the loss-cone not by two-body encounters, but by other mechanisms: the
tidal separation of stellar or compact binaries approaching the SMBH \citep{FS2018},
accretion-disk-mediated TDEs \citep{KMS2016}, TDEs produced by a recoiling SMBH \citep{GM2008,LLB2012} or by a merging SMBH binary \citep{2016NatSR...635629H,LLBS2017}. Finally, we note that star clusters possessing radially biased velocity distribution, especially at high binding energy, or having nonspherical gravitational potential may help to increase the rate of TDEs with $e<1$ (not too close to 1) and extend the $\beta$ distribution to $\beta>1$ in the empty loss-cone regime; see also the detailed discussion at the end of Section~\ref{sec:cons}.
%

%
%

%
%

\section{Conclusion}
\label{sec:cons}

Understanding TDEs and their light curves provides a clue to the properties of the central SMBH, the accretion disk around it, and to the stellar density and velocity distributions in the nuclear star cluster surrounding the SMBH. The link between TDEs and central star clusters in
galactic nuclei has been a classical subject of seminal papers, such as
\cite{FR1976,BW1976} finding the classical density distribution near central SMBHs. \citet{1977ZhETF..73.1587D,1977PAZh....3..295D} first noted that accretion and the tidal disruption of stars with low angular momentum cause the density profile to flatten out toward the black hole (the energy distribution function $f(E)$ drops towards zero as
they showed; today we would call this the empty loss-cone region---see also \citet{1978Ap&SS..59..171O} for a summary of the topic at the time). 
CK78 put this on a more quantitative footing by using the technique of solving the orbit-averaged Fokker--Planck equation. Among the classical work in this field, also Rees's conjecture about the fate of tidal debris \citep{R1988} is most noteworthy; it has been expanded more recently by \cite{HZL2018} and \cite{2020ApJ...900....3P}, looking for critical eccentricities that separate partial from full mass loss (hyperbolic) and partial from full mass accretion (eccentric).

Our study generalizes and expands this by computing approximate distributions of bound and unbound stars by predicting analytically (and comparing with $N$-body data) the number densities
as a function of eccentricity $e$ and penetration factor $\beta$, both of which are key parameters for the prediction of the observational appearance of TDEs.

By following the generalized model of CK78 for bound and unbound stars, we predict that tidally disrupted stars, for all penetration factors, occupy only a small range in eccentricities, much smaller than the critical eccentricities cited above. We estimate some minimum, maximum, and typical eccentricities for tidally disrupted stars. They are all very close to the parabolic case, either very marginally eccentric or very marginally hyperbolic (see Fig.~\ref{fig_boundary}).

\cite{2001MNRAS.327..995A} and \cite{2004MNRAS.352..655A} proposed another model for the loss-cone in a spherical star cluster. It is interesting to note that they have also derived the density and velocity dispersion of bound and unbound loss-cone stars by using moment
equations of the basic Fokker--Planck equation and very similar principles to CK78 and this paper. Due to the use of moment equations (the so-called gaseous model of star clusters; see, e.g., \citealt{GS1994}) their analysis is completely based on density and velocity
profiles rather than orbits with energy and angular momentum (or eccentricity and penetration factor). Their model takes into account
an anisotropic velocity distribution also for the unbound stars. In the future, a more quantitative
comparison of the two models could be done.

Our theoretical predictions have all been tested against the data of our previously published direct $N$-body simulations \citep{HZL2018}; comparison with our analytical model helps to understand the scaling behavior of the $N$-body simulations, since we can still not yet do realistic particle numbers for them. Our primary conclusions are
summarized as follows:

\begin{enumerate}
\item
Our results provide the number density of bound tidally disrupted stars as a function of orbital eccentricity $e$ and penetration factor $\beta$; in practice, we used the cumulative numbers of disrupted stars $N_{\rm TD}(e)$ and $N_{\rm TD}(\beta)$ over some simulated time, since they can be directly compared with simulation results.
To get them, we fit the stellar density with double power law profile and solved the corresponding velocity dispersion via the Jeans equation. We also use double-power-law functions to model the energy distribution of tidally disrupted stars (obtained from the $N$-body data).
\item
From these results, we have analytically derived three characteristic orbital eccentricities: $e_{\rm min}$, $e_{\rm max}$, and $e_{\rm lcb}$ in the loss-cone region, where $e_{\rm min}$ and $e_{\rm max}$ take the minimum and maximum values for a given $\beta$, respectively, whereas $e_{\rm lcb}$ represents the orbital eccentricity which gives the boundary between the empty and full loss-cone regimes. These eccentricities are given by equations (\ref{Eq-b-boundary}), (\ref{Eq-u-boundary}), and (\ref{Eq-Q=1}), respectively. We have confirmed that the stars causing TDEs are distributed between $e_{\rm min}$ and $e_{\rm max}$ on the $e-\beta$ plane by N-body experiments. Moreover, we find most of the bound stars are focused between $e_{\rm lcb}$ and $e=1$, i.e., in the full loss-cone regime, whereas the remaining bound stars are originating from the empty loss-cone regime. This result is consistent with the loss-cone theory. 
\item 
We conclude from the limiting eccentricity values that they are very close to the parabolic case, and far away from the critical eccentricities for complete debris accretion or complete debris escape from the SMBH. We have shown that this conclusion holds also for larger more realistic black hole masses.
\end{enumerate}

Our model of angular momentum diffusion at a given energy value $E$, as given in Eq,~\ref{eq:appq}, uses a free parameter 
$Q_{\rm boost}$ for fitting to our simulation results. \cite{Merritt2013} suggested using the steady-state solution of a Fokker-Planck equation in angular momentum space, for every energy value, in order to obtain the flux across the loss-cone boundary, i.e. a value of $Q_{\rm boost}$ in our terminology. This concept is based on the assumption that steady state in angular momentum space is achieved much faster than in energy space; it is used by the \texttt{PhaseFlow} 1D Fokker-Planck code \citep{Vasiliev2017}.
In our paper, we prefer not to follow such a two-timescale approach, rather keep all our fitting procedures in energy space and use the free factor $Q_{\rm boost}$. There may be several factors which could affect the clean separation of angular momentum and energy diffusion time scales. Most notable are rotation, axisymmetric gravitational potentials---see our earlier 2D Fokker-Planck model with full 2D representation of angular momentum diffusion near the loss-cone in \cite{Fiestas2010,Fiestas2012}, based on the 2D Fokker-Planck code used by \cite{Einsel1999}, but also strong anisotropy could have an impact here \cite{Szolgyen2019}.
Arguably the use of \texttt{PhaseFlow} will be a good method to
quickly get the $n(e)$ and $n(\beta)$ distributions for TDEs in galaxy models, using energy distributions, and directly aimed at real systems \citep{Pfister2019,Pfister2020,Bortolas2023}.
In any case, from the knowledge of $n(e)$ and $n(\beta)$ one could estimate the distribution of peak mass fallback rate in different galaxies, since the peak mass fallback rate depends on both $\beta$ \citep{GRR2013} and $e$ \citep{HSL2013,2020ApJ...900....3P}.

Some issues remain to be subject of further work;
in our $N$-body simulations, we have adopted a fixed position and mass of the central black hole. For a star cluster with an intermediate mass black hole (IMBH), for example, the ratio of black hole mass to stellar mass will be smaller than in this paper, and the Brownian motion of the IMBH will not be suppressed. The Brownian motion of the black hole modifies the energy distribution of stars so that the number density and $e$-$\beta$ distributions can be significantly affected, which is subject of our future work.

Our model is a high-resolution direct $N$-body model of a nuclear star cluster, following individual stellar orbits and TDEs. We have measured the distribution of orbital parameters of tidally disrupted stars and compared the results with a semi-analytical model. Our $N$-body models are not restricted to spherical symmetry, even though in this paper we do study only spherical nuclear star clusters. We do not intend to predict detailed TDE rates for specific galaxies, such as other models based on 1D Fokker-Planck theory 
\citep[see, e.g.,][]{SM2016,Pfister2019,Pfister2020,Bortolas2023}. Rather, we are interested in the analysis of the stellar distribution, relaxation and accretion processes only in the inner zone of a nuclear star cluster (stellar mass limited to ten times the black hole mass). Models based on 1D Fokker-Planck theory are computationally much faster and can extend much farther out, but rely on approximations such as spherical symmetry and a steady state in angular momentum diffusion.

Future work in the domain of our $N$-body simulation model is to
include a stellar mass spectrum, stellar populations with different ages, direct stellar collisions and relativistic dynamics of stellar mass black holes in the nuclear star cluster. This goes along with a more realistic treatment of tidal disruptions (partial and full; \citealt{Zhongetal2022,McLeodetal2013}) as well as direct plunges and relativistic or dissipative inspirals \citep[see, e.g., our recent paper][]{LZB2023}. Furthermore the assumption of spherical symmetry will be relaxed in favour of rotating, axisymmetric \citep{Fiestas2010,ZBS2015}  and triaxial models \citep{NormanSilk1983,PM2002,PM2004,MP2004}. All of these will disturb the steady-state picture underlying our current paper and have interesting consequences for TDEs and produce also gravitational-wave events instead of TDEs. 

\pagebreak

\begin{acknowledgments}
S.Z. has been supported by the National Natural Science Foundation of China (NSFC~11603067) and acknowledges the support from Yunnan Astronomical Observatories, Chinese Academic of Sciences. K.H. has been supported by the Korea Astronomy and Space Science Institute (KASI) under the R\&D program supervised by the Ministry of Science, ICT and Future Planning, and by the Basic Science Research Program through the National Research Foundation of Korea (NRF) funded by the Ministry of Education (2016R1A5A1013277, 2017R1D1A1B03028580, and 2020R1A2C1007219 (K.H.)). K.H. has been supported by the National Supercomputing Center with supercomputing resources including technical support (KSC-2019-CRE-0082 (K.H.)) and in part by the National Science Foundation under Grant No.~NSF PHY-1748958. K.H. has been also financially supported during the research year of Chungbuk National University in 2021. The authors acknowledge the Yukawa Institute for Theoretical Physics (YITP) at Kyoto University. Discussions during the YITP workshop YITP-T-19-07 on International Molecule-type Workshop ``Tidal Disruption Events: General Relativistic Transients" were useful to complete this work.
The authors also acknowledge support by the Chinese Academy of Sciences (CAS) through the Silk Road Project at NAOC. We are grateful for the support from the Sino-German Center (DFG/NSFC) under grant no. GZ1289.
S.L., P.B. and R.S. acknowledge the Strategic Priority Research Program (Pilot B) Multi-wavelength gravitational wave universe of the Chinese Academy of Sciences (No.~XDB23040100). S.L. and R.S. acknowledges Yunnan Academician Workstation of Wang Jingxiu (No.~202005AF150025). The work of PB was supported by the Volkswagen Foundation under the special stipend No.~9D154.
PB thanks the support from the special program of the Polish Academy of Sciences and the U.S. National Academy of Sciences under the Long-term program to support Ukrainian research teams grant No.~PAN.BFB.S.BWZ.329.022.2023.

\end{acknowledgments}

\bibliographystyle{aasjournal}
\bibliography{TD-2019-rev}

\appendix


\section{Details of the variable transformations}
\label{Appendix-relation}

In this work we use the unusual variables $e$ and $\beta$ as independent phase space variables,
instead of the more widely used standard variable pairs $(\mathcal{R},E)$ or $(J,E)$. We present here the
variable transformations used in the main text, as well as the corresponding Jacobian determinants.

For the star \emph{bound} to the SMBH, the specific angular momentum, $J$, is given by
\begin{equation}
J = \sqrt{G M_{\rm BH}a(1-e^2)},
\label{Eq-J}
\end{equation}
\noindent
where $e<1$ and $a$ is the semimajor axis of the star. Putting $e=0$ into
the above equation, we get the circular angular momentum $J_{\rm c}=\sqrt{G M_{\rm BH} a}$.
At the $e=1$ limit,
\begin{equation}
J \approx \sqrt{2 G M_{\rm BH} r_{\rm p} },
\label{Eq-J_rp}
\end{equation}
\noindent
where $r_{\rm p} = a(1-e)$. Equating $r_{\rm p}$ with $r_{\rm t}$,
we get the loss-cone angular momentum:
\begin{equation}
J_{\rm lc} = \sqrt{2 G M_{\rm BH} r_{\rm t}}.
\label{Eq-J_lc}
\end{equation}
\noindent
Substituting equation (\ref{Eq-J_rp}) into equation (\ref{eq:angmom}),
we obtain the relation between the original independent variable $\mathcal{R}$
and the new independent variables $(e,E)$,

\begin{equation}
\mathcal{R}(e,E) = 1-e^2.
\label{Eq-R-e}
\end{equation}

Substituting equation (\ref{Eq-J_lc})
into equation (\ref{eq:angmom}), we obtain
\begin{equation}
\mathcal{R}_{\rm lc} = \frac{2 r_{\rm t}}{a} = \frac{2|E|}{|E_{\rm t}|}.
\label{Eq-R_lc}
\end{equation}
\noindent
where $E_{\rm t}=-GM_{\rm BH}/(2 r_{\rm t})$ is the specific energy
estimated at the tidal disruption radius.

From equations~(\ref{Eq-J_rp}) and (\ref{Eq-J_lc}),
$r_{\rm t}$ and $r_{\rm p}$ can be expressed as
\begin{equation}
r_{\rm p} = \frac{J^2}{2 G M_{\rm BH}},
\label{Eq:r_p-J2}
\end{equation}
and
\begin{equation}
r_{\rm t} = \frac{J_{\rm lc}^2}{2 G M_{\rm BH}},
\label{Eq:r_t-Jlc2}
\end{equation}
respectively. Through equation~(\ref{eq:angmom}) and $\beta=r_{\rm t}/r_{\rm p}$, we obtain the relation between the original independent variable $\mathcal{R}$ and the new independent variables $(\beta,E)$,
\begin{equation}
\mathcal{R}(\beta,E) = \frac{|E|}{|E_{\rm t}|}\frac{2}{\beta},
\label{Eq-beta}
\end{equation}

The original independent variables $(\mathcal{R},E)$ expressed with the new
independent variables $(\beta,e)$ are given by
$\mathcal{R}(\beta,e) = 1-e^2$ and $E(\beta,e) = \beta(1-e)E_{\rm t}$, respectively.
For the reader's convenience, we also derive the variable
$\mathcal{R}$ expressed with $(J,E)$ in the bound case,
which is $\mathcal{R}(J,E) = 2|E|J^2 /(GM_{\rm BH})^2$.

For an \emph{unbound} star, we have
$r_{\rm p} = -a(e\!-\!1) = a (1-e)$ with $a<0,r_{\rm p}>0$.
\footnote{In the simulation, we record the position $\mathbf{r}$ and velocity $\mathbf{v}$ at the time when a star enters $r_{\rm t}$. The two-body eccentricity of a unbound star is $e=\sqrt{2EJ^2/(GM_{\rm BH})^2+1}$, where $J=|\mathbf{r}\times\mathbf{v}|$ and $E=|\mathbf{v}|^2/2-GM_{\rm BH}/r$. Then $\beta$ is computed with equation~\ref{Eq-beta-e-unbound}.}
The relation between the original independent variable $\beta$
and the new independent variables $(e,E)$ is

\begin{equation}
\beta(e,E) = \frac{E}{|E_{\rm t}|(e\!-\!1)},
\label{Eq-beta-e-unbound}
\end{equation}
through $E= - GM/(2a) > 0$.

The Jacobian determinants corresponding to the above variable transformations are summarized in
Table~\ref{table-Jacob}.

\begin{table}[htbp]
\begin{center}
\caption{Table of the variable transformation and the associated Jacobian determinant
\label{table-Jacob}}
\begin{tabular}{cccc}
  \tableline
  Orbit type & Original variable set & New variable set  &   Jacobian determinant   \\
  \hline
  bound  & $(\mathcal{R},E)$ & $(e,E)$  & $2e$ \\
  bound  & $(\mathcal{R},E)$ & $(\beta,E)$ & $2|E|/(|E_{\rm t}|\beta^2)$ \\
  bound  & $(\mathcal{R},E)$ & $(\beta,e)$ & $2e(1\!-\!e)|E_{\rm t}|$ \\
  bound  & $(\mathcal{R},E)$ & $(J,E)$ & $4|E|J/(GM_{\rm BH})^2$ \\
  \hline
  unbound & $(\beta,E)$ & $(e,E)$ &  $E/[|E_{\rm t}|(e\!-\!1)^2]$\\
  \tableline
\end{tabular}
\tablecomments{The Jacobian determinants take a positive sign. }
\end{center}
\end{table}

\end{document}